\documentclass[12pt]{iopart}
\usepackage{amssymb}
\usepackage{xspace}
\usepackage{graphicx}
\usepackage{appendix}
\usepackage{cite}
\begin{document}
\title[Phase Dependent Forcing and Synchronization]{Phase Dependent Forcing and Synchronization in the three-sphere model of {\em Chlamydomonas}}
\author{Rachel R. Bennett and Ramin Golestanian}
\address{Rudolf Peierls Center for Theoretical Physics, University of Oxford, Oxford, OX1 3NP, UK}
\ead{ramin.golestanian@physics.ox.ac.uk}
\begin{abstract}
The green alga {\it Chlamydomonas} swims with synchronized beating of its two flagella, and is experimentally observed to exhibit run-and-tumble behaviour similar to bacteria. Recently we studied a simple hydrodynamic three-sphere model of {\it Chlamydomonas} with a phase dependent driving force which can produce run-and-tumble behaviour when intrinsic noise is added, due to the non-linear mechanics of the system. Here, we consider the noiseless case and explore numerically the parameter space in the driving force profiles, which determine whether or not the synchronized state evolves from a given initial condition, as well as the stability of the synchronized state. We find that phase dependent forcing, or a beat pattern, is necessary for stable synchronization in the geometry we work with.
\end{abstract}
\pacs{05.45.45.Xt, 87.16.Qp, 47.63.-b}
\maketitle

\begin{section}{Introduction}
Microorganisms swim in the low Reynolds number regime where viscous forces dominate, inertia is negligible and the familiar propulsion methods of larger organisms become ineffective \cite{purcell:1976,hb:1965,lp:2009,childress:1981}. Fluid flow is governed by the Stokes equation, which is time reversible. A necessary condition on a periodic swimming stroke in order to achieve net propulsion is that it is non-time reversible \cite{sw:1987}. Inspired by sperm cells, which achieve propulsion by propagation of bending waves through their flagellum, Taylor demonstrated that propulsion is possible in a viscous environment by studying the propagation of waves on an infinite sheet \cite{taylor:1951}. Purcell showed that a swimmer needs at least two compact degrees of freedom to break the time reversal symmetry and achieve net propulsion \cite{purcell:1976}.

Many microorganisms swim using flagella \cite{lighthill:1976}; there are two fundamentally different types of flagella: bacterial flagella and eukaryotic flagella (or cilia). Eukaryotic flagella form bends when microtubules on one side of the flagella `walk' or `slide' along the microtubules on the other side \cite{ajlrrw:2008}. The propagation of bends allows the flagella to form beat patterns that can break the time reversal symmetry. For example, the first half of an individual cilium's beat cycle, called the {\em power stroke}, has the cilium sticking out and pushing the fluid, while the second half, called the {\em recovery stroke}, has the cilium bent as it returns to its original position \cite{bs:1974}.

Our understanding of propulsion at low Reynolds number has been developed by theoretical model microswimmers. Lighthill demonstrated a model that can achieve net propulsion by studying periodic shape deformations of a nearly spherical swimmer, showing that the swimming velocity is at most of the order of the square of the amplitude of the deformations \cite{lighthill:1952}. Purcell's three-link swimmer was studied by Becker {\it et al.}, who determined the swimming direction and velocity for different angle amplitudes and relative link lengths \cite{purcell:1976,bks:2002}. A useful one-dimensional model  is the linear three-sphere swimmer, where three beads are connected by two rods that change length with a non-reciprocal pattern \cite{ng:2004}. Dreyfus {\it et al.} studied a rotational analogue of the three-sphere swimmer \cite{dbs:2005}. Avron {\it et al.} presented a  more efficient swimmer consisting of a pair of bladders which exchange their volume and vary the distance between them \cite{ako:2005}. There have been several experimental realisations of artificial low Reynolds number swimmers \cite{dbrfsb:2005,tgps:2008,lkbcc:2009}.

When two sperms swim close to each other, their tails beat in synchrony \cite{gray:1928} and Taylor studied this using his waving sheet model with hydrodynamic interactions \cite{taylor:1951}. Coordinated beating of flagella or cilia is important for a range of processes including motility, efficient pumping of fluid and symmetry breaking in developing embryos \cite{gray:1928,gl:1999,berg:2004,ov:2011,nywigmh:2005}. Theoretical and experimental models have been studied to show that synchronization can occur through hydrodynamic interaction and that it is relevant to bacterial swimming and pumping by arrays \cite{cbj:2002,cjb:2003,kbvbp:2003,gyu:2011,MKim:2004,Reich:2005,kn:2006,vj:2006,lr:2006,gj:2007,nel:2008,qjgbp:2009,el:2009,ug:2010,gu:2010,klblc:2010,ug:2011,ws:2011,bkdlc:2012,dbkvoo:2012,lbckc:2012,ug:2012}. 
Flagellar synchronization is observed in {\it Chlamydomonas}, a unicellular green alga that swims using two flagella that beat with a breaststroke pattern \cite{ringo:1967,rn:1985,rn:1998}. The cell has diameter $\sim10{\rm \mu m}$ and swims with velocity $~100{\rm \mu m/s}$ so the Reynolds number is ${\rm Re} \sim 10^{-3}$ and inertia is negligible. During normal swimming, the flagella beat in synchrony. These periods of synchrony are interrupted by periods of asynchronous beating and during these asynchronies, there is a large change in the cell's orientation \cite{ptdgg:2009,gpt:2009}. This is analogous to run-and-tumble behaviour observed in bacteria.

Simple models have helped us understand better the intricacies of low Reynolds number swimming and hydrodynamic synchronization, and a recent development has been to combine these two effects in the context of a simple three-sphere model for the swimming of {\it Chlamydomonas} \cite{fj:2012,bg:2012,pf:2012}. This simple model captures some of the important features of {\it Chlamydomonas}, namely, the ability to swim, the exact role of hydrodynamic interactions \cite{fj:2012,pf:2012}, the existence of stable synchronized states and an emergent run-and-tumble behaviour which is observed when we add white noise to the driving force \cite{bg:2012}. Here, we consider the model without added noise and explore its phase diagram and full parameter space to see when the model evolves into the synchronized state. We also investigate the stability of the synchronized state under various conditions, and the different types of behaviour that can be obtained form the model.
These studies have revealed a number of intriguing features.

\end{section}

\begin{section}{The Model}
We arrange three beads in the $x - y$ plane, each of radius $a$, on a frictionless scaffold, as shown in figure \ref{fig:model}.
\begin{figure}
   \begin{center}
   \includegraphics[width=0.7\columnwidth]{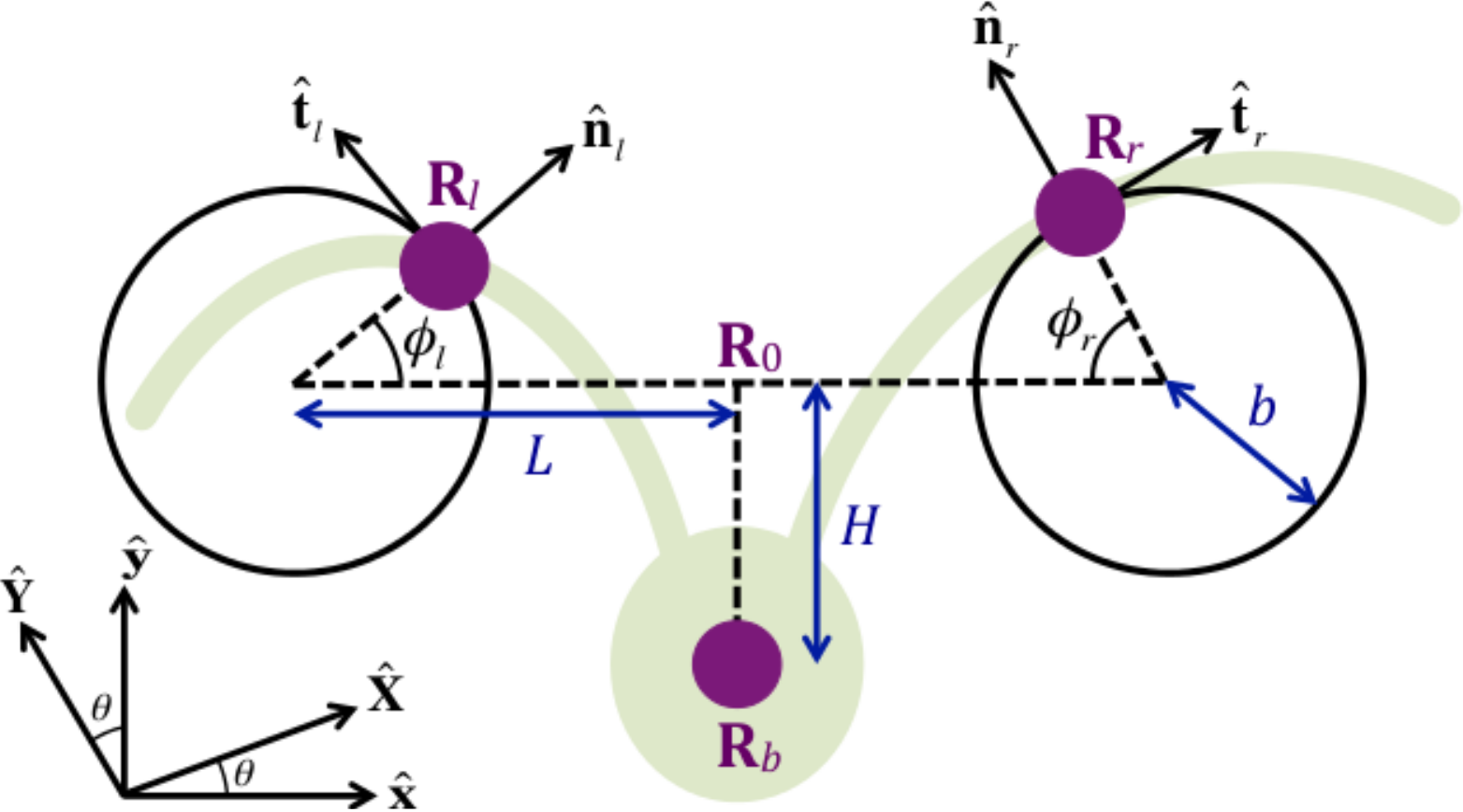}
   \end{center}
   \caption{A three sphere model of \textit{Chlamydomonas}. The left and right beads represent the flagella and move on circular trajectories in the cell frame. The back bead represents the cell body. The scaffold is shown to define the reference frame of the cell which has origin $\mathbf{R}_0$ in a lab frame, but it does not interact with the fluid. The green underlay is a schematic of a \textit{Chlamydomonas} cell.}
   \label{fig:model}
\end{figure}
We refer to the left, right and back beads with the subscripts `$l$', `$r$' and `$b$', respectively. Let $\mathbf{R}_0$ be the origin of the cell reference frame with respect to a lab frame. The cell axes $\hat{\bf x},\hat{\bf y}$ make an angle $\theta(t)$ with the lab axes $\hat{\bf X},\hat{\bf Y}$. The left and right beads model the flagella and move on circular trajectories in the cell frame of radius $b$ in opposite directions and with phases $\phi_l$ and $\phi_r$; the back bead models the cell body and is fixed with respect to the cell frame. The positions and velocities of the beads are
\begin{eqnarray}
   & \mathbf{R}_l=\mathbf{R}_0-L\mathbf{\hat{x}}+b\mathbf{\hat{n}}_l, \qquad
   & \mathbf{\dot{R}}_l=\mathbf{\dot{R}}_0+L\dot{\theta}\mathbf{\hat{y}}+b(\dot{\phi}_l-\dot{\theta})\mathbf{\hat{t}}_l, \label{eq:Rl} \\
   & \mathbf{R}_r=\mathbf{R}_0+L\mathbf{\hat{x}}+b\mathbf{\hat{n}}_r, \qquad
   & \mathbf{\dot{R}}_l=\mathbf{\dot{R}}_0-L\dot{\theta}\mathbf{\hat{y}}+b(\dot{\phi}_r+\dot{\theta})\mathbf{\hat{t}}_r, \label{eq:Rr} \\
   & \mathbf{R}_b=\mathbf{R}_0-H\mathbf{\hat{y}},
   & \mathbf{\dot{R}}_b=\mathbf{\dot{R}}_0-H\dot{\theta}\mathbf{\hat{x}}. \label{eq:Rb}
\end{eqnarray}
where the unit vectors $\mathbf{\hat{n}}_i$ and $\mathbf{\hat{t}}_i$ are in the normal and tangent directions of the circular trajectory of $\mathbf{R}_i$. The left and right beads are driven by tangential forces $F_l^t$ and $F_r^t$ respectively. Normal forces $F_l^n$ and $F_r^n$ are exerted by the beads in order to be constrained to the circular trajectories. The force on the back bead is such that the swimmer is force free and torque free:
\begin{equation}
   \mathbf{F}_l+\mathbf{F}_r+\mathbf{F}_b=0, \qquad
   \mathbf{T}_l+\mathbf{T}_r+\mathbf{T}_b=0,
   \label{eq:forcetorquefree}
\end{equation}
where $\mathbf{F}_i=F_i^t\mathbf{\hat{t}}_i+F_i^n\mathbf{\hat{n}}_i$ for $i=l,r$ and $\mathbf{T}_j=\mathbf{R}_j\times \mathbf{F}_j$ for $j=b,l,r$.

The forces and velocities are related through hydrodynamic interactions between the beads:
 \begin{eqnarray}
   & \mathbf{\dot{R}}_l=\frac{1}{\xi}\mathbf{F}_l+(\mathbf{G}_{lr}-\mathbf{G}_{lb})\cdot \mathbf{F}_r-\mathbf{G}_{lb}\cdot \mathbf{F}_l, \label{eq:Rld} \\
   & \mathbf{\dot{R}}_r=\frac{1}{\xi}\mathbf{F}_r+(\mathbf{G}_{rl}-\mathbf{G}_{rb})\cdot \mathbf{F}_l-\mathbf{G}_{rb}\cdot \mathbf{F}_r, \label{eq:Rrd} \\
   & \mathbf{\dot{R}}_b=-\frac{1}{\xi}(\mathbf{F}_l+\mathbf{F}_r)+\mathbf{G}_{bl}\cdot \mathbf{F}_l+\mathbf{G}_{br}\cdot \mathbf{F}_r, \label{eq:Rbd}
\end{eqnarray}
where $\xi=6\pi \eta a$ is the friction coefficient of each bead ($\eta$ is viscosity of the ambient fluid). In the limit when $a$ is small compared with all other length scales, the hydrodynamic interaction is described by the Oseen tensor $\mathbf{G}_{ij}=\frac{1}{8 \pi \eta |\mathbf{r}_{ij}|}(\mathbf{I}+\hat{\mathbf{r}}_{ij}\hat{\mathbf{r}}_{ij}) $ with $\mathbf{r}_{ij}=\mathbf{r}_i-\mathbf{r}_j$ \cite{oseen:1927}.

The phase difference $\delta = \phi_r-\phi_l$ evolves according to
\begin{eqnarray}
   \dot{\delta} = & \frac{1}{b} \Bigg[(\dot{\mathbf{R}}_r-\dot{\mathbf{R}}_b)\cdot \hat{\mathbf{t}}_r-(\dot{\mathbf{R}}_l-\dot{\mathbf{R}}_b)\cdot \hat{\mathbf{t}}_l  +\nonumber \\
   & 2 \dot{\theta} \Big(\cos{(\delta/2)}\big(L\cos{\phi}-H\sin{\phi}\big)-b\Big) \Bigg],
   \label{eq:deltadot}
\end{eqnarray}
where $\phi=(\phi_l+\phi_r)/2$ and
\begin{eqnarray}
   \dot{\theta} =& \frac{1}{2} \Big[\frac{(\dot{\mathbf{R}}_l-\dot{\mathbf{R}}_b)\cdot \hat{\mathbf{n}}_l}{H \cos{(\phi-\delta/2)}+L\sin{(\phi-\delta/2)}}-\nonumber \\
   & \frac{(\dot{\mathbf{R}}_r-\dot{\mathbf{R}}_b)\cdot \hat{\mathbf{n}}_r}{H \cos{(\phi+\delta/2)}+L\sin{(\phi+\delta/2)}}\Big].
   \label{eq:thetadot}
\end{eqnarray}
We solve equation \ref{eq:deltadot} numerically for a choice of stroke pattern (driving forces) $F_i^t(\phi_i)$, $i=l,r$. The long term behaviour of $\delta$ depends on the stroke pattern and in many cases the initial condition. We compute $\dot{\delta}, \dot{\theta}$ and $\dot{\phi}$ to leading order in $a/L$, since we do not require hydrodynamic interactions for synchronization \cite{fj:2012}, but we include the next order hydrodynamic term when computing the velocities, since we need this second order affect to achieve a net swimming velocity. 

\end{section}

\begin{section}{Swimming velocity in the synchronized state}
First we consider the synchronized state where $F_l^t(\phi)=F_r^t(\phi)$ and  $\delta=0$, so that $\phi_l=\phi_r=\phi$. We do not worry about the stability of the synchronized state, which we consider in the next section, and assume that the swimmer stays in this state. Since the Reynolds number is low, we need to ask `does the model achieve net propulsion?' If hydrodynamic interactions are not  included, then the cell just moves forwards and backwards and there is no net motion. However, if we include hydrodynamic interactions, which vary in strength around the cycle, then the symmetry in the swimming stroke is broken and net propulsion is achieved.

The magnitude and direction of the net swimming velocity depends on the ratios $H/b$ and $L/b$. Figure \ref{fig:HLvel} shows the net swimming velocity in the ${\bf \hat{y}}$ direction for a range of $(H/b,L/b)$ and constant driving force $F^t(\phi) = F_0$. For $L\gtrsim2.25 |H|$, swimming is in the positive direction, otherwise the cell swims in the negative direction. Polotzek and Friedrich in reference \cite{pf:2012} give the following explanation of why the cell may swim in either direction. The instantaneous velocity $v=f/g$ is the ratio of the force $-f {\bf \hat{y}}$, which has to be applied to the back bead to prevent it from moving, and the friction coefficient $g$ associated with towing the swimmer in the ${\bf \hat{y}}$ direction. The force $f$ oscillates during a stroke cycle and the hydrodynamic interactions, which reduce the magnitude of $f$, are strongest when the beads are closest together. On the other hand, the friction coefficient is largest when the beads are furthest apart and smallest when the beads are close together. The geometry of the swimmer determines which effect dominates and therefore whether the net swimming is in the positive or negative direction.
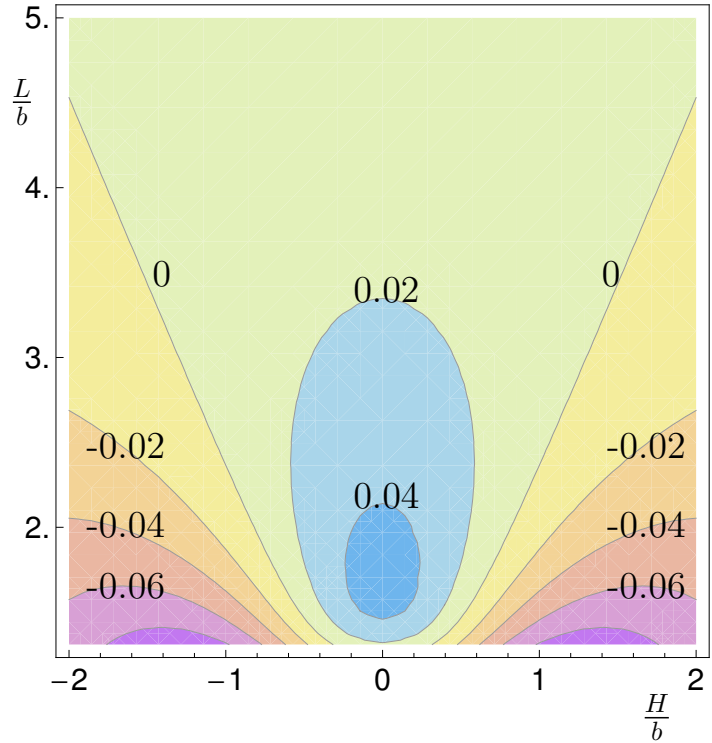
\begin{figure}
   \begin{center}
   \includegraphics[width=0.58\columnwidth]{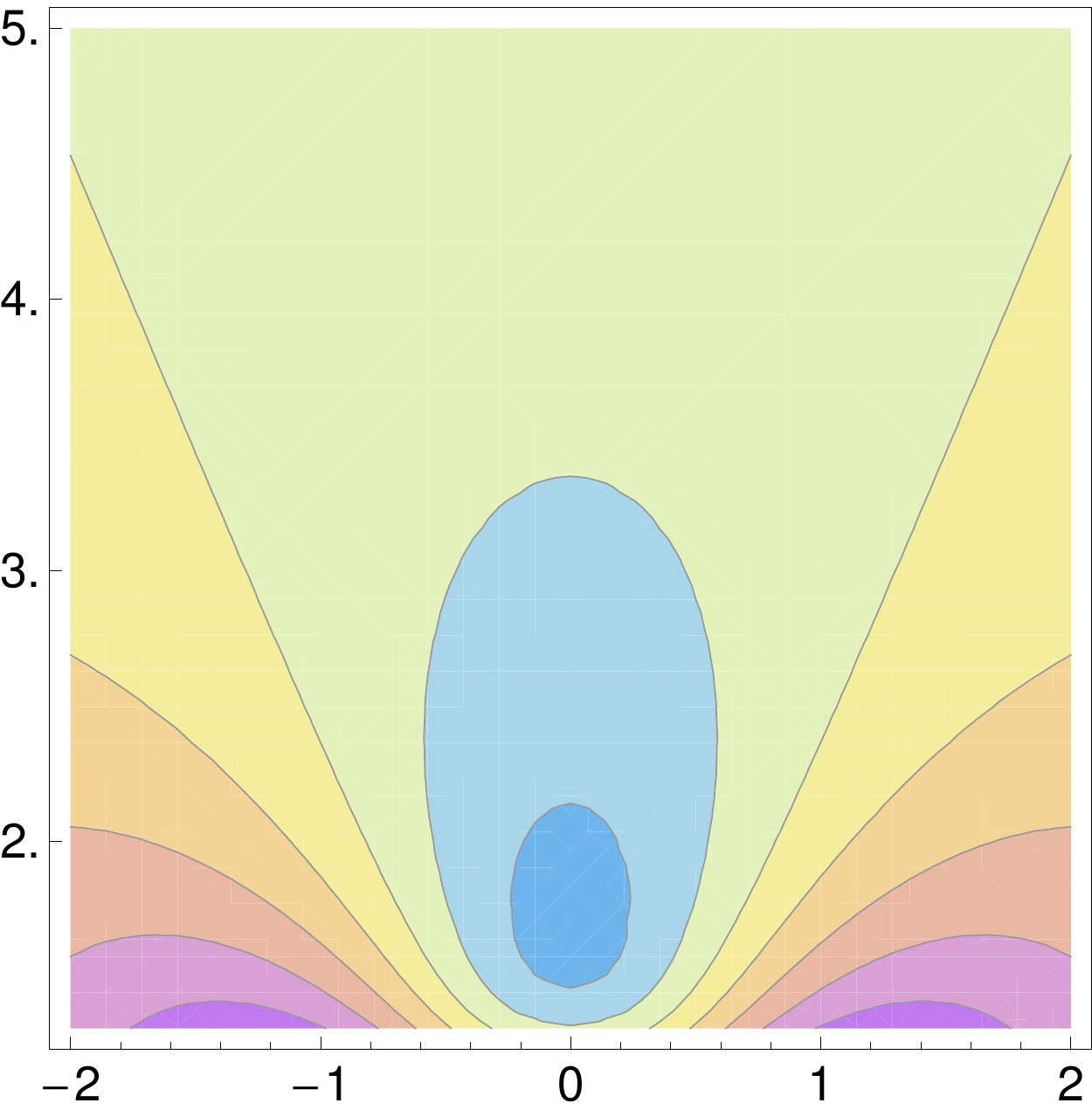} 
   \end{center}
   \begin{picture}(10,10)(0,0)
   \put(327,13){\large{$\frac{H}{b}$}} 
   \put(88,246){\large{$\frac{L}{b}$}}
   \put(142,180){\large{0}}
   \put(312,180){\large{0}}
   \put(218,174){\large{0.02}}
   \put(218,96){\large{0.04}}
   \put(117,115){\large{-0.02}}
   \put(314,115){\large{-0.02}}
   \put(117,85){\large{-0.04}}
   \put(314,85){\large{-0.04}}
   \put(117,62){\large{-0.06}}
   \put(314,62){\large{-0.06}}
   \end{picture}
   \caption{Contour plot of dependence of velocity on the parameter lengths. The contours show the velocities in units of $F_0/(6\pi \eta b)$. The zero contour lies approximately on the line $L=2.25|H|$ for sufficiently large $H/b$.}
   \label{fig:HLvel}
\end{figure}
Herein we fix the values $a/L = 1/33, H/b = 1/5$ and $L/b = 2$, which results in forwards swimming for $F_0>0$.

The the net velocity is also affected by the force profile and we consider driving forces $F^t(\phi)$ such that $\int_0^{2\pi} {\rm d}\phi F^t(\phi)  =2\pi F_0 $, where $F_0$ is a fixed average force. The net velocity $\bar{v}$ can be written as $\bar{v}=1/T\int_0^T {\rm d}t \dot{R_b}$, where $\dot{R_b}={\bf \dot{R}}_b\cdot {\bf \hat{y}}$, $T=\int_0^{2\pi}{\rm d}\phi/\dot{\phi}$ is the period, and we can write $\int_0^T {\rm d}t \dot{R_b}= \int_0^{2\pi}{\rm d}\phi \dot{R_b}(\phi)/\dot{\phi}(\phi)$. In the synchronized state the force dependence cancels in the ratio $\dot{R_b}/\dot{\phi}$, so the force only enters the net velocity expression through the $1/T$ term. In order to maximise the net velocity, we must minimise the period $T = \int_0^{2\pi} {\rm d}\phi/\dot{\phi}(\phi)$, where we can write $\dot{\phi}(\phi) = F^t(\phi)\Phi(\phi)$. Minimising $T$ with the constraint $\int_0^{2\pi} {\rm d}\phi F^t(\phi)  =2\pi F_0 $ tells us that a constant force profile $F^t(\phi)=F_0$ maximises the net velocity. Clearly, increasing $F_0$ increases the net velocity. However as we shall see in the next section, the synchronized state is not stable when we choose a constant force profile. Friedrich {\it et al.} showed that a constant driving force can give a stable synchronized state if we change the direction of rotation of the beads, so $F_0<0$ \cite{fj:2012,pf:2012}, which is equivalent to changing the sign of $H$, but here we choose to work with $F_0>0$ and $H>0$.

\end{section}

\begin{section}{Synchronization and stability}
We consider force profiles of the forms $F_i^t(\phi_i)=F_0\big(1+a_i^{(n)} \cos{n \phi_i}\big)$ and $F_i^t(\phi_i) = F_0\big(1+b_i^{(n)}\sin{n\phi_i}\big)$, where $-1<a_i^{(n)},b_i^{(n)}<1$ and $i=l,r$. The definition of the synchronized state that we use here is zero (or integer multiple of $2\pi$) phase difference between the two flagella, i.e. when $\delta = 2\pi n$, $n \in \mathbb{Z}$. Initially we tried to analyst the the synchronization stability by linear stability analysis, but we are unable to do this because we cannot perform a valid Taylor expansion when $\phi=\phi_s$, where $H\cos{\phi_s}+L\sin{\phi_s}=0$. We work numerically to avoid this Taylor expansion; for further details see the appendix.

We identify five main types of stability of the synchronized state by looking at the evolution of $\delta$ from different initial conditions $\delta(\phi_0)=0.1$ for a number of $\phi_0$ equally spaced in the range $\phi_0\in[0,2\pi)$: (i) All the initial conditions $\delta(\phi_0)=0.1$ evolve into the synchronized state for all $\phi_0$ (the synchronized state is stable). (ii) Some choices of $\phi_0$ evolve into the synchronized state and others choices evolve into an oscillating state, but there is a larger number of $\phi_0$ that lead to synchronization than the number of $\phi_0$ that leads to oscillations. (iii) Some choices of $\phi_0$ evolve into the synchronized state and others choices evolve into an oscillating state; the numbers of $\phi_0$ that lead to each type of behaviour are similar. (iv) Some choices of $\phi_0$ evolve into the synchronized state and others choices evolve into an oscillating state, but there is a larger number of $\phi_0$ that lead to oscillations than the number of $\phi_0$ that leads to synchronization. (v) All choices of $\phi_0$ evolve into the oscillating state, (the synchronized state is unstable).

Although the choice of initial condition $\delta_0=0.1$ is arbitrary, we want to know how likely it is that a small perturbation from the synchronized state will decay back to synchronization, or whether it is likely to evolve into an oscillating state. This choice of $\delta_0$ is suitable for this purpose. For type (v) stability, if we start in the synchronized state, then it is likely that some numerical noise will kick $\delta$ into an oscillating state. It is possible for a small amount of noise to kick the synchronized state into an oscillating state for types (ii), (iii), (iv), with low probability for type (ii), then increasing probability for type (iii) and then type (iv).

\subsection{Equal beat patterns}
First we consider the case where $F_l^t(\phi)=F_r^t(\phi)=F^t(\phi)$. For each choice of coefficient and initial condition, $\delta$ evolves either to an integer multiple of $2\pi$ and remains at this value (synchronization); or it reaches a state where it oscillates about $\pi$ sinusoidally; or it reaches a periodic state near zero, but never reaching zero. Figure \ref{fig:devolutionS1} shows examples of these three cases and the corresponding orientation $\theta$.
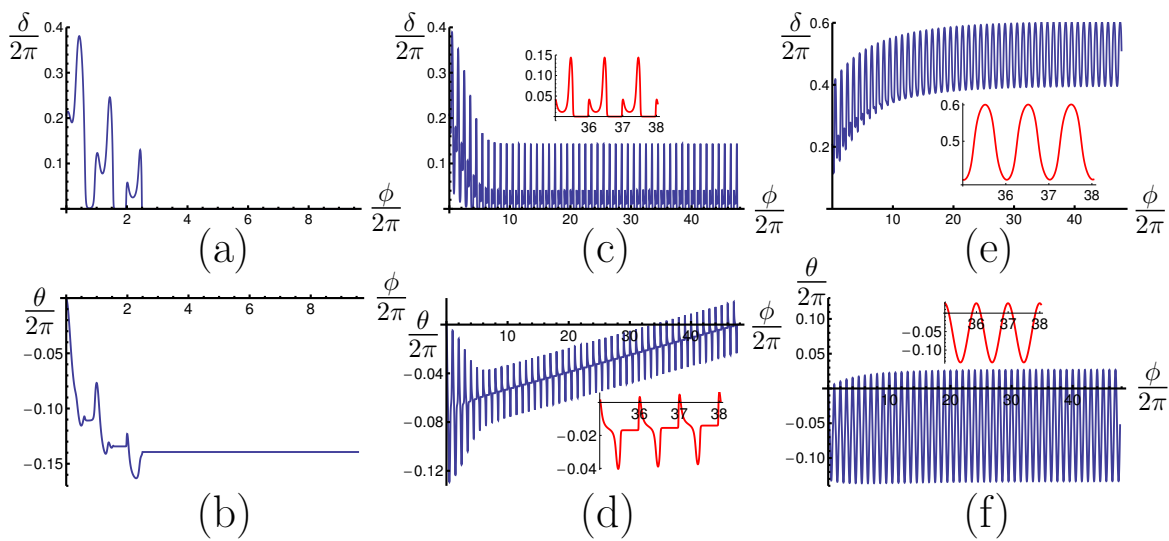
\begin{figure}
   \begin{center}
   \hspace{1.6mm}
   \includegraphics[width=0.27\columnwidth]{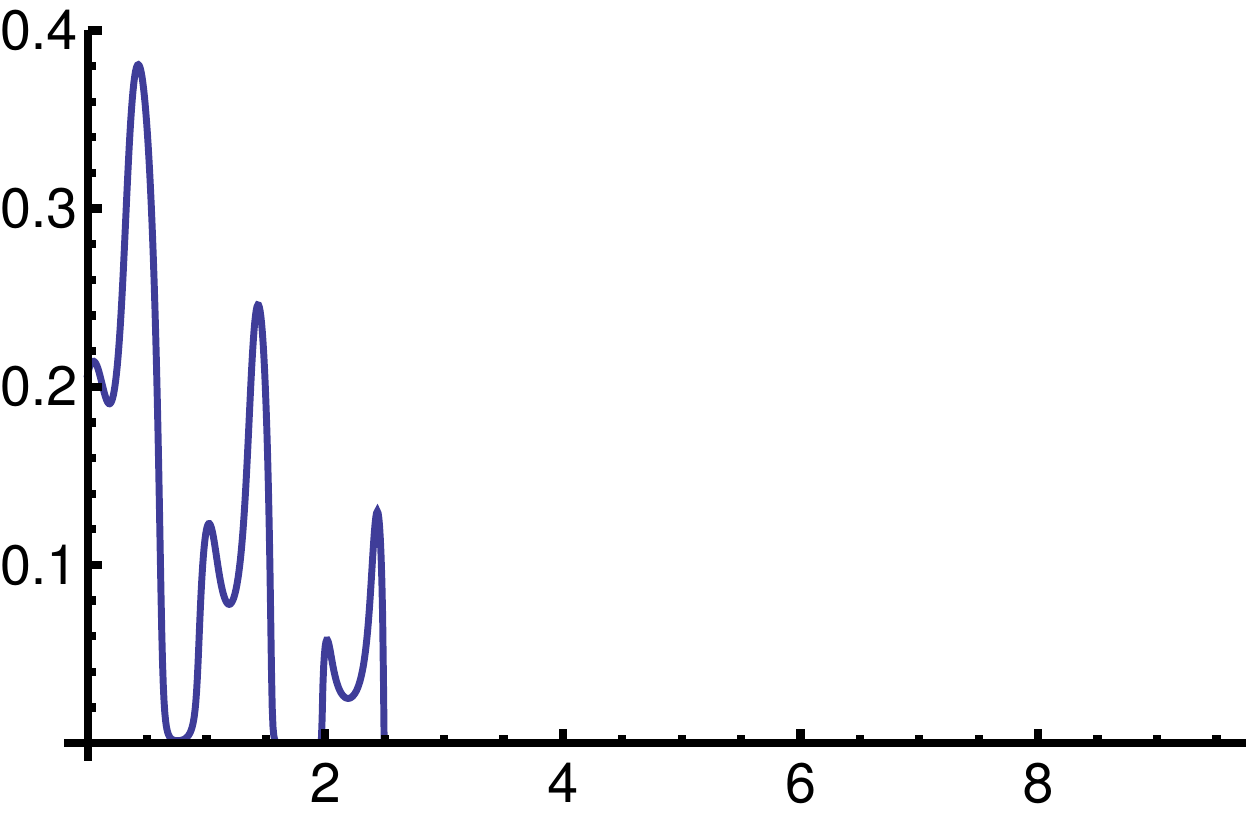}
   \hspace{5.8mm}
   \includegraphics[width=0.27\columnwidth]{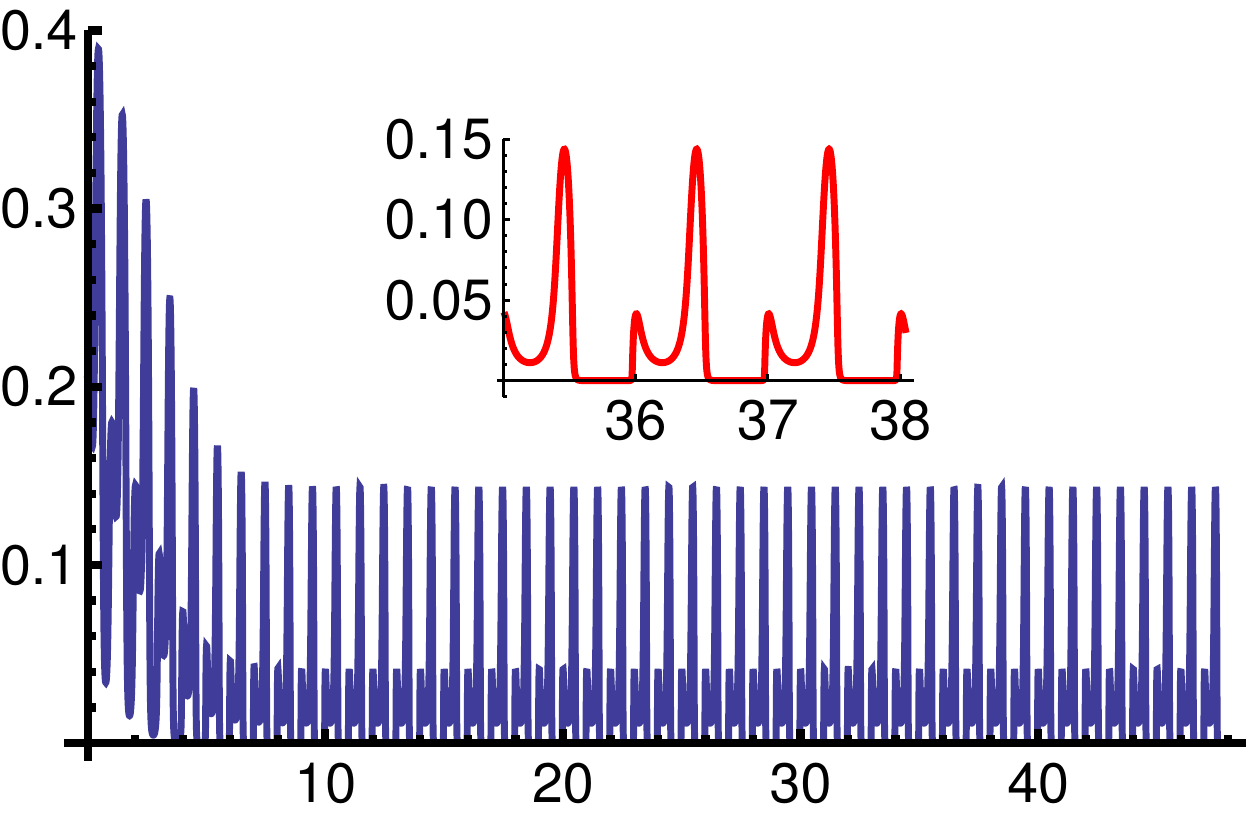}
   \hspace{5.8mm}
   \includegraphics[width=0.27\columnwidth]{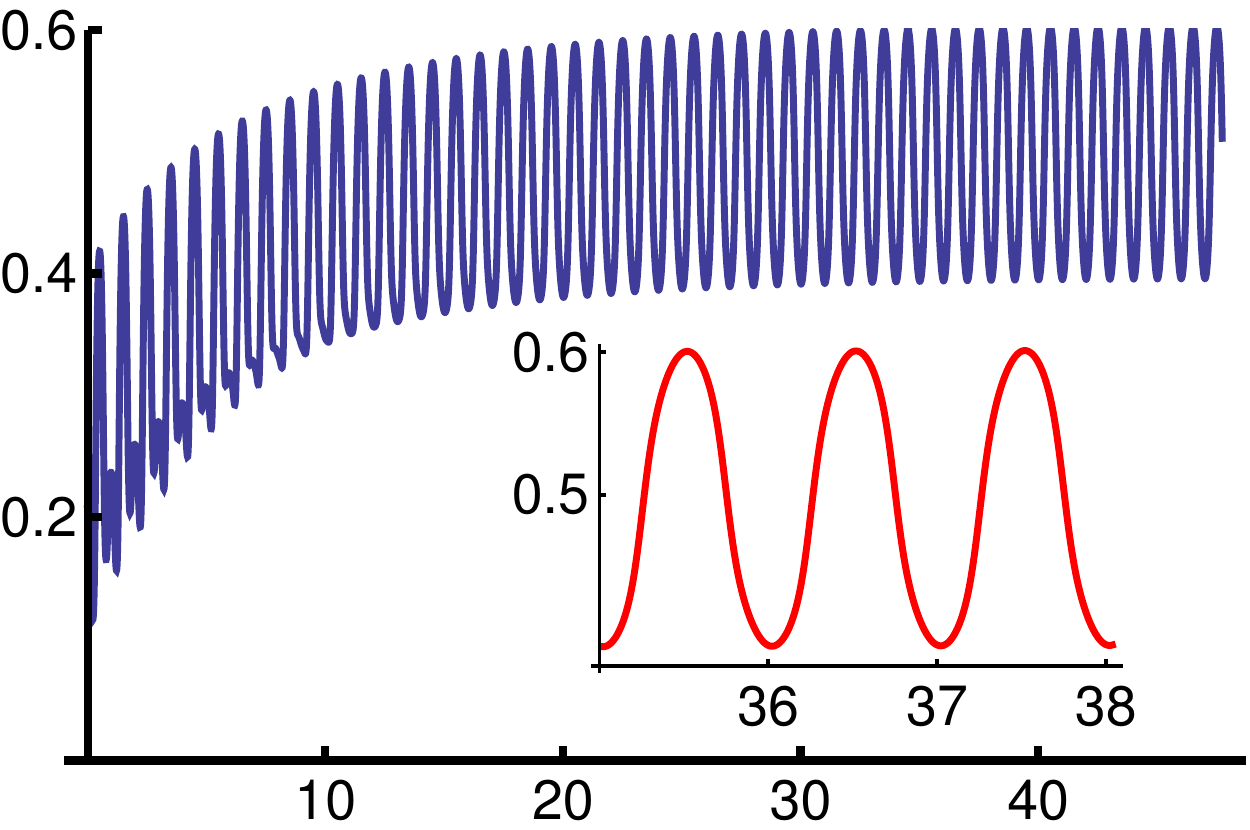} \\
   \vspace{8.0mm}
   \includegraphics[width=0.29\columnwidth]{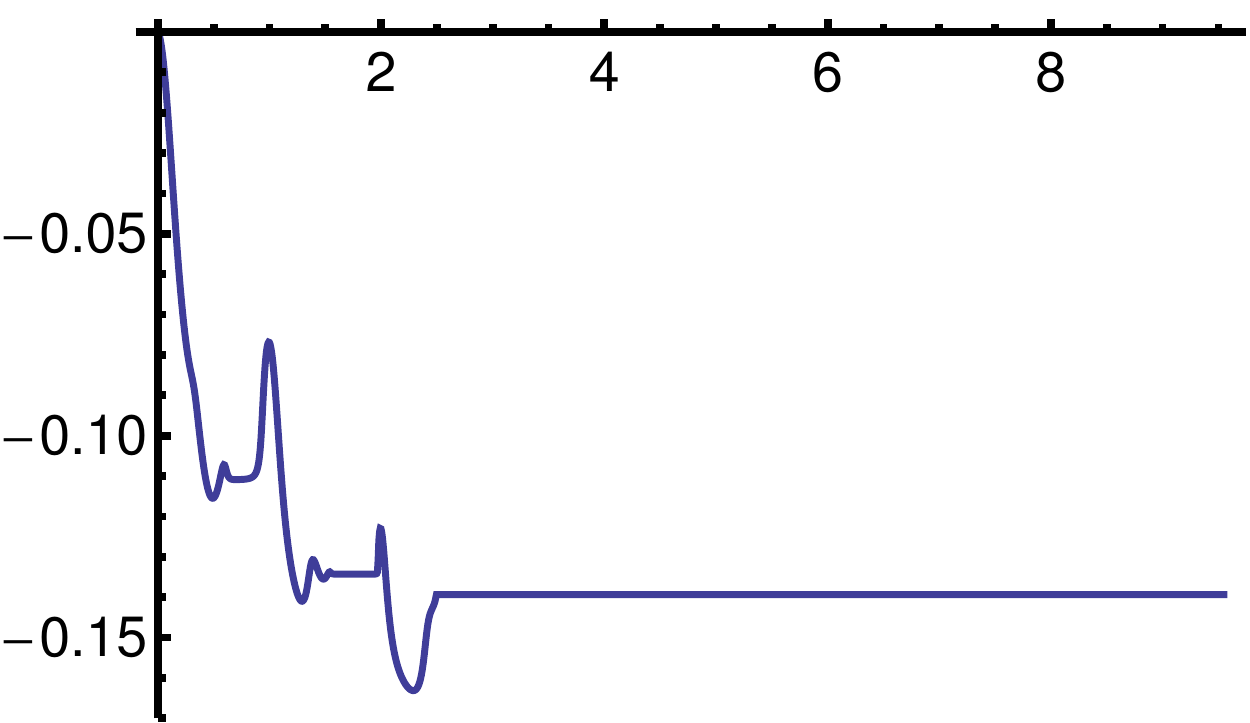}
   \hspace{2.3mm}
   \includegraphics[width=0.29\columnwidth]{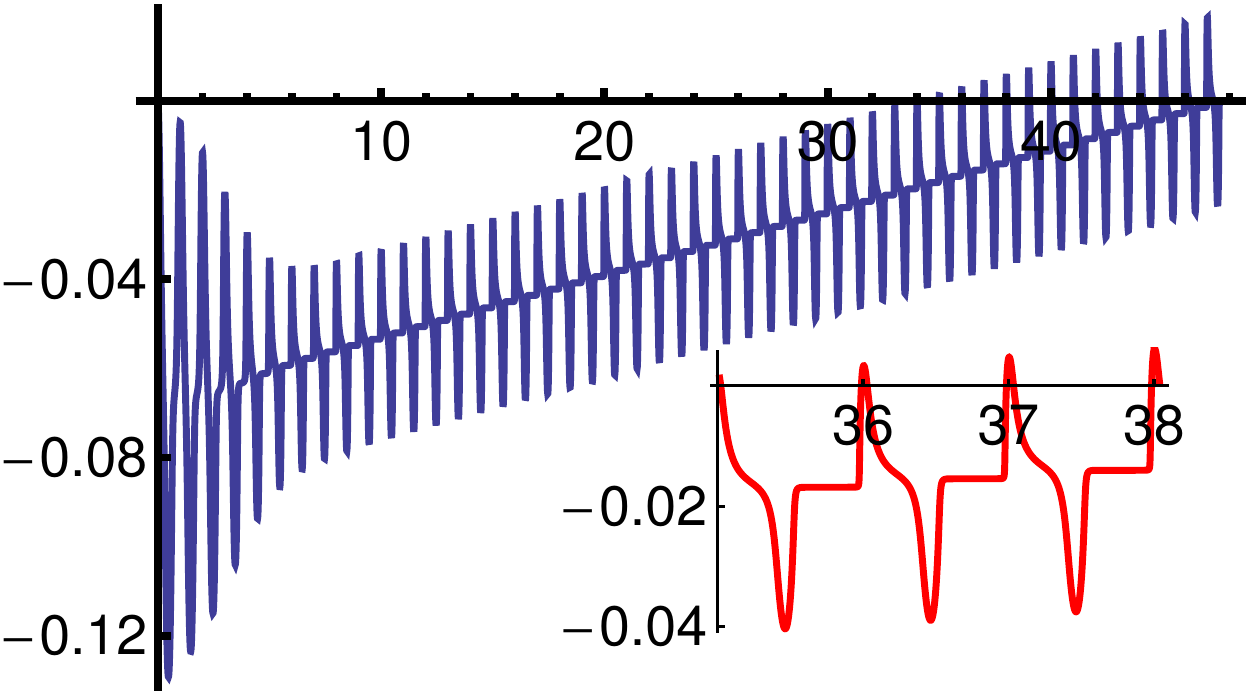}
   \hspace{2.5mm}
   \includegraphics[width=0.29\columnwidth]{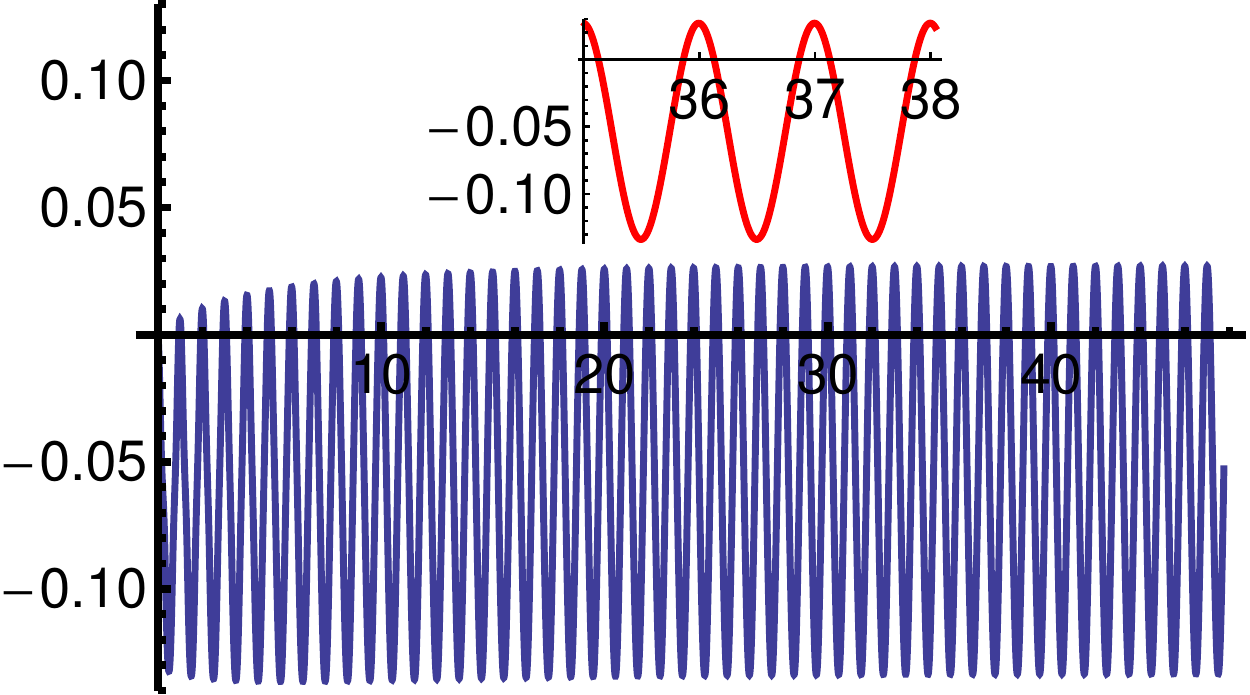}
   \end{center}
   \begin{picture}(10,10)(0,0)
   \put(145,127){\large{$\frac{\phi}{2\pi}$}}
   \put(290,126){\large{$\frac{\phi}{2\pi}$}}
   \put(434,126){\large{$\frac{\phi}{2\pi}$}}

   \put(8,192){\large{$\frac{\delta}{2 \pi}$}}
   \put(154,192){\large{$\frac{\delta}{2 \pi}$}}
   \put(300,192){\large{$\frac{\delta}{2 \pi}$}}

   \put(14,88){\large{$\frac{\theta}{2 \pi}$}}
   \put(147,93){\large{$\frac{\phi}{2\pi}$}}

   \put(159,79){\large{$\frac{\theta}{2 \pi}$}}
   \put(289,82){\large{$\frac{\phi}{2\pi}$}}

   \put(306,99){\large{$\frac{\theta}{2 \pi}$}}
   \put(435,59){\large{$\frac{\phi}{2\pi}$}}

   \put(80,110){\Large{(a)}}
   \put(80,10){\Large{(b)}}
   \put(225,110){\Large{(c)}}
   \put(225,10){\Large{(d)}}
   \put(370,110){\Large{(e)}}
   \put(370,10){\Large{(f)}}
   \end{picture}
   \caption{Evolution of $\delta$ for driving force $F^t(\phi)=F_0 (1+b^{(1)} \sin{\phi})$ and initial condition $\delta(\phi=0)=1.3$ with (a) $b^{(1)}=0.7$, (b) $b^{(1)}=0.4$, (c) $b^{(1)}=-0.1$. The insets show a small part of the plot in more detail. The bottom axis $\phi/2\pi$ shows the number of cycles which increases monotonically with time. Bottom row: corresponding $\theta(\phi)$.}
   \label{fig:devolutionS1}
\end{figure}
When $\delta$ oscillates about $\pi$, then the orientation oscillates about some fixed value. The cycle averaged motion is in a straight line, but the cell jiggles from side to side as well as backwards and forwards as it moves along. When $b^{(1)}=0.4$ and the oscillations are near zero, there is a net drift in the orientation so the net motion of the cell is along a curved trajectory.

For many choices of $F^t(\phi)$, the initial condition determines whether $\delta$ evolves into the synchronized state or the oscillating state. Figure \ref{fig:ICphasediagram} shows the the dependence of $\delta$ evolution on initial condition for $F^t(\phi)=F_0 (1+a^{(1)} \cos{\phi})$. A 63 $\times$ 69 grid is shown where each square represents an initial condition $(\phi_0,\delta_0)$. A black square represents an initial condition for which $\delta \to 0$ after a sufficiently long time; a white square represents an initial condition for which $\delta$ continues to oscillate periodically as $t \to \infty$.
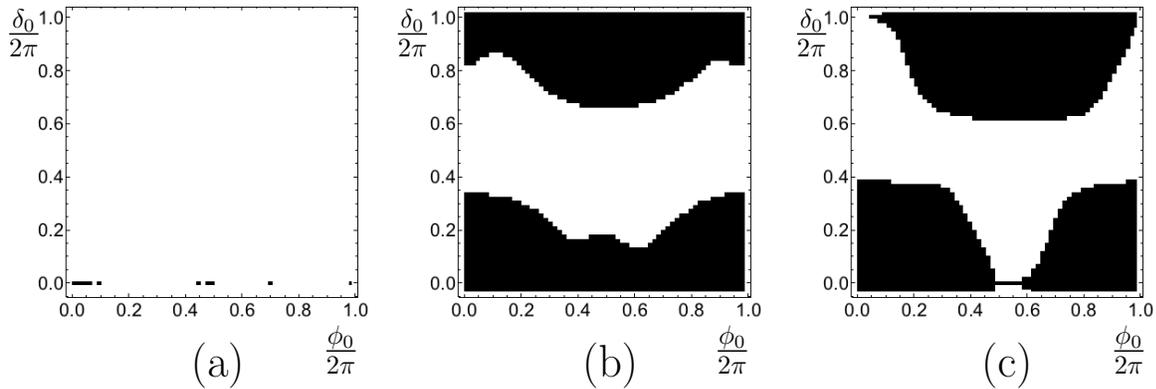
\begin{figure}
   \begin{center}
   \includegraphics[width=0.28\columnwidth]{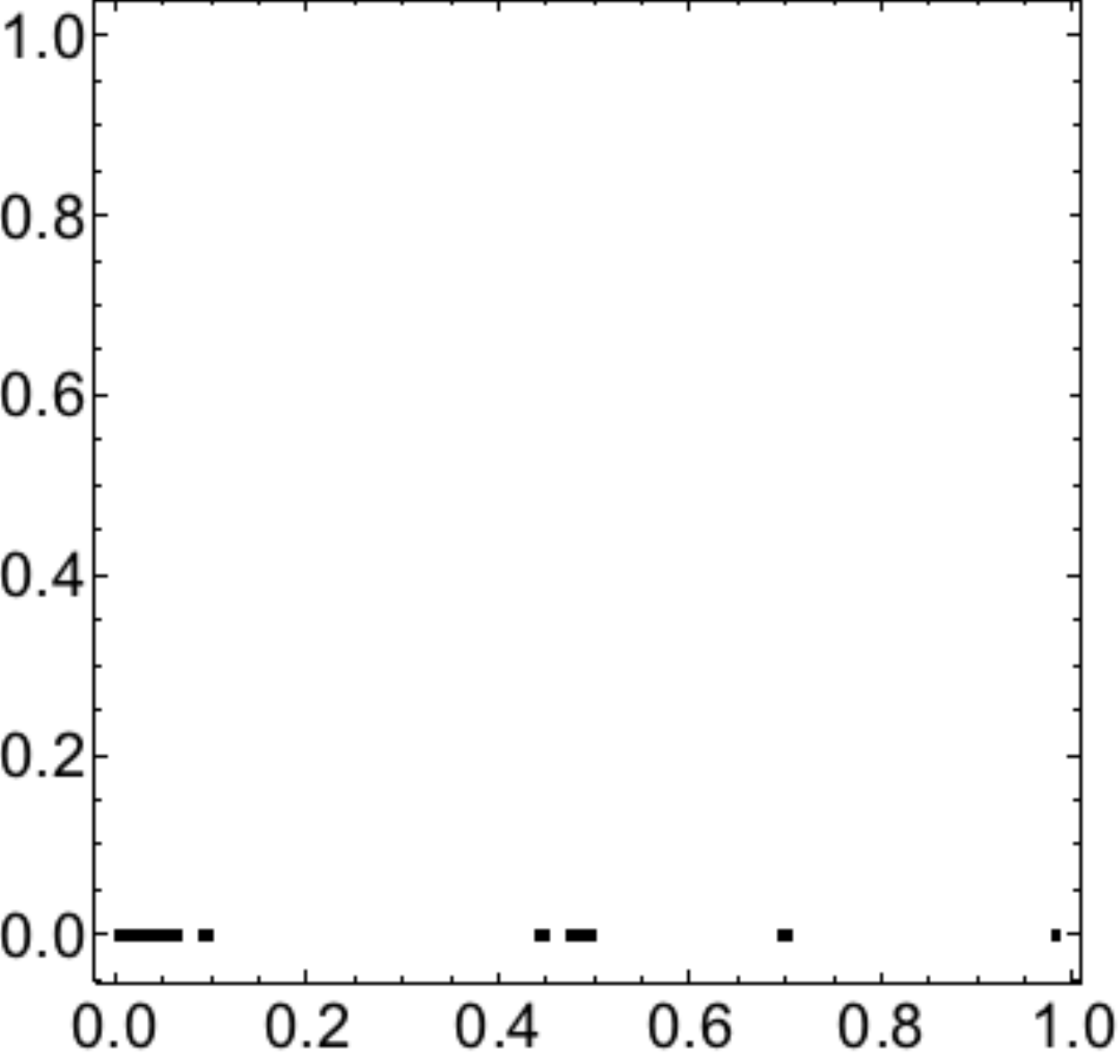}
   \hspace{5.5mm}
   \includegraphics[width=0.28\columnwidth]{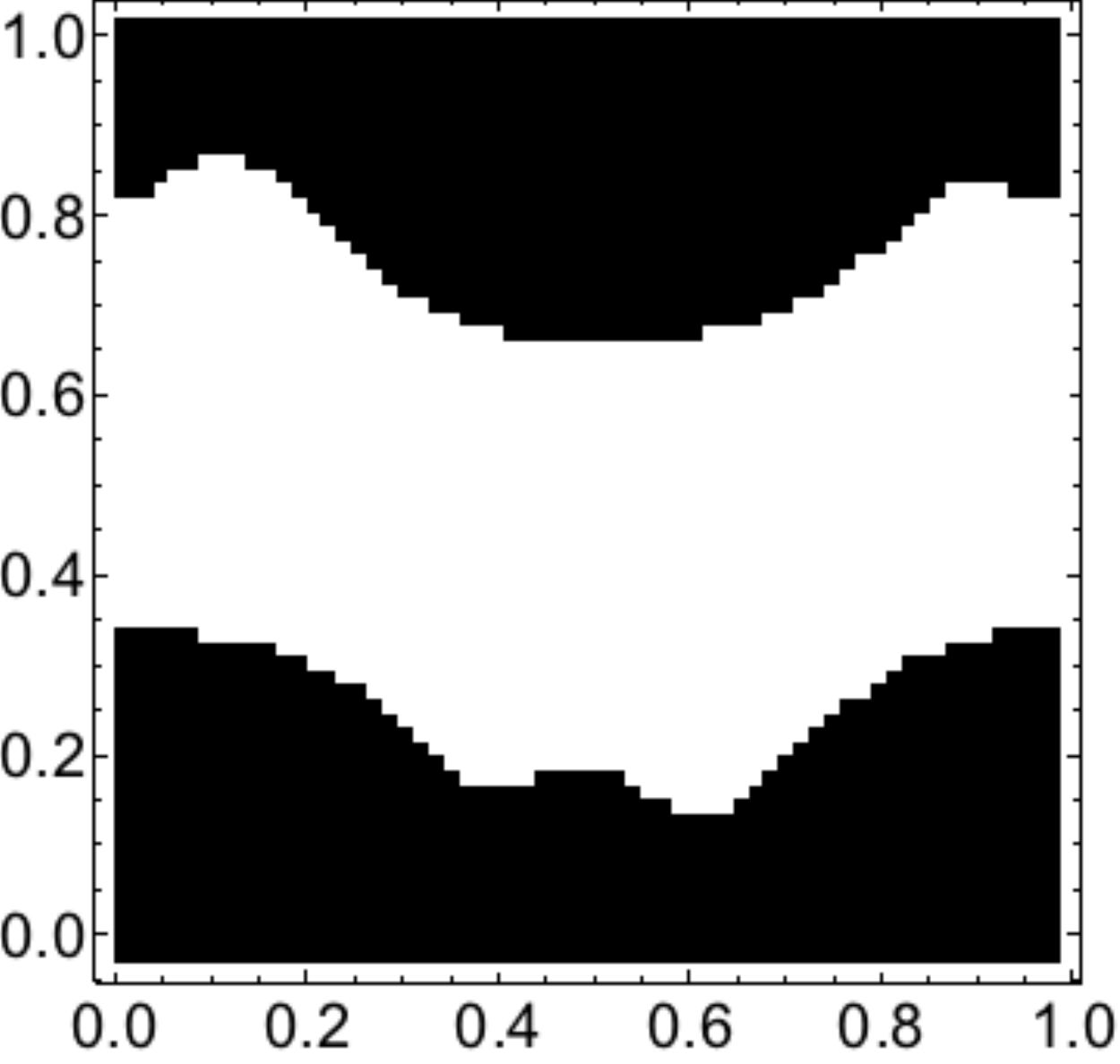}
   \hspace{5.5mm}
   \includegraphics[width=0.28\columnwidth]{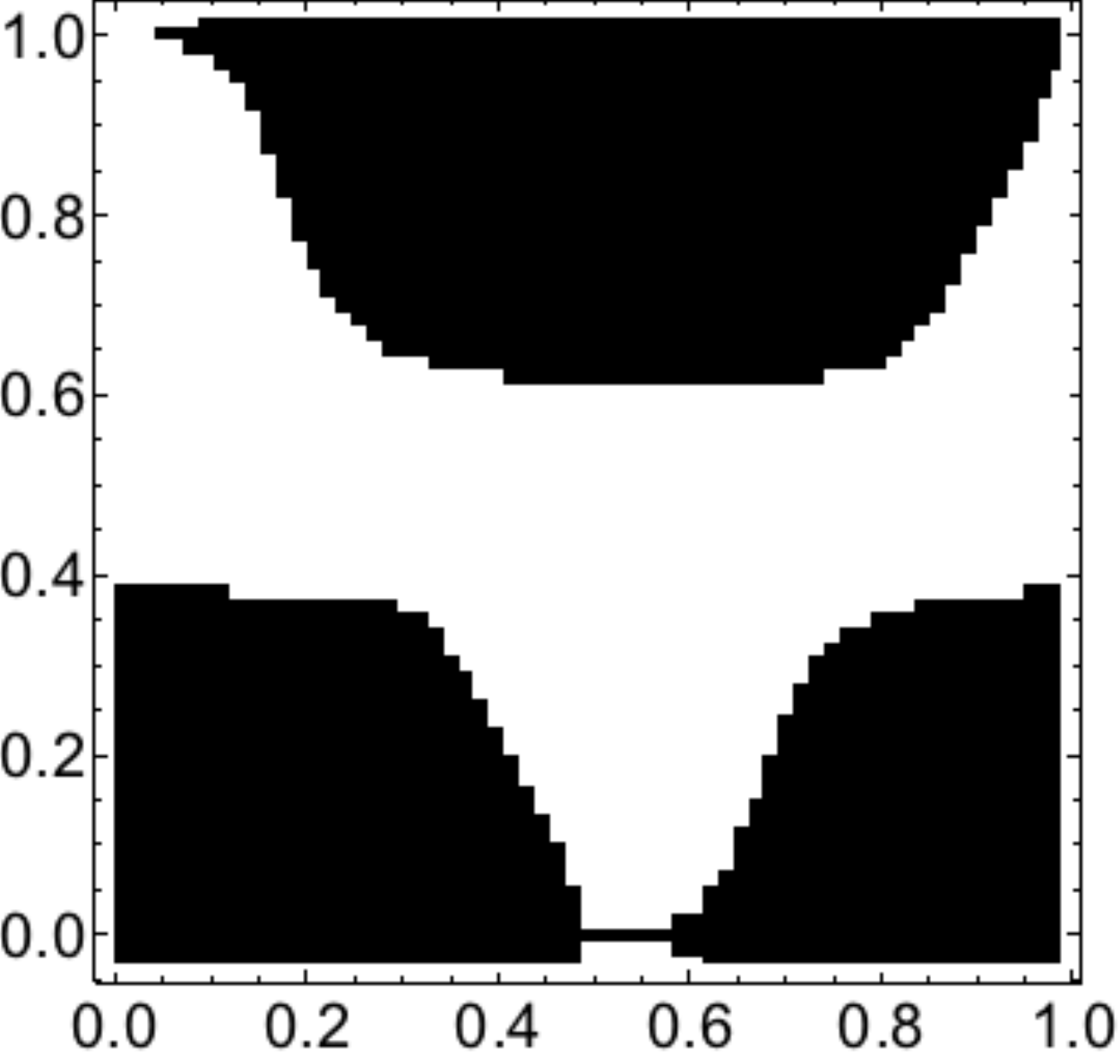}
   \end{center}
   \begin{picture}(10,10)(0,0)

   \put(120,10){\large{$\frac{\phi_0}{2\pi}$}}
   \put(0,130){\large{$\frac{\delta_0}{2\pi}$}}

   \put(268,10){\large{$\frac{\phi_0}{2\pi}$}}
   \put(147,130){\large{$\frac{\delta_0}{2\pi}$}}

   \put(418,10){\large{$\frac{\phi_0}{2\pi}$}}
   \put(298,130){\large{$\frac{\delta_0}{2\pi}$}}

   \put(70,4){\Large{(a)}}
   \put(218,4){\Large{(b)}}
   \put(369,4){\Large{(c)}}

   \end{picture}
   \caption{Phase diagrams showing how $\delta$ evolves for different initial conditions for $F^t(\phi)=1+a^{(1)} \cos{\phi}$ with (a) $a^{(1)}=0.4$, (b) $a^{(1)}=0.6$, (c) $a^{(1)}=0.8$. Black squares represent initial conditions which lead to the synchronized state and white squares represent initial conditions which lead to a periodic oscillating state. The stability of the synchronized state is: (a) type (v) unstable; (b) type (i) stable; (c) type (ii).}
   \label{fig:ICphasediagram}
\end{figure}

For $a^{(1)}=0.4$, all initial conditions lead to an oscillating state and the synchronized state is unstable (type (v)). The black squares in figure \ref{fig:ICphasediagram} are initially in the synchronized state. Many squares along the line $\delta=0$ are white because a small amount of numerical noise drives the system away from the synchronized state. For $a^{(1)}=0.6$, initial conditions close to $\delta=0$ lead to the synchronized state, but initial conditions far from $\delta=0$ lead to an oscillating state. The synchronized state is stable (type (i)). For $a^{(1)}=0.8$, most initial conditions close to $\delta=0$ lead to the synchronized state, but a few initial conditions close to $\delta=0$ lead to an oscillating state and we have type (ii) stability; if $\delta$ starts close to the synchronized state, it is likely to evolve into the synchronized state and it will stay in the synchronized state if there is no noise, but it is also possible for the cell to start close to the synchronized state and move away into an oscillating state. In an oscillating state the cell can still swim, but there will be more side to side movement. There appear to be a few white squares on the line $\delta_0 = 2\pi$, however this is because the grid does not lie exactly on the $2\pi$ line, the grid contains points $\delta=6.2$ and $\delta = 6.3$ and this small deviation from the synchronized state $\delta=2\pi$ is enough for evolution into an oscillating state for a few choices of $\phi_0$.

Similar behaviour is observed for other harmonics, but with different ranges of coefficients giving the different stability types for the synchronized state. For example, figure \ref{fig:harm4} shows the 4th harmonic for three different coefficients of cosine and initial condition $\delta(\phi=0)=0.1$. Figure \ref{fig:harm4}(a) shows that $\delta$ oscillate about $\pi$ for $a^{(4)}=0.4$, oscillations are close to zero for $a^{(4)}=0.8$ (it is interesting to note the 4 peaks in every cycle), and $\delta$ evolves into the synchronized state for $a^{(4)}=0.9$.
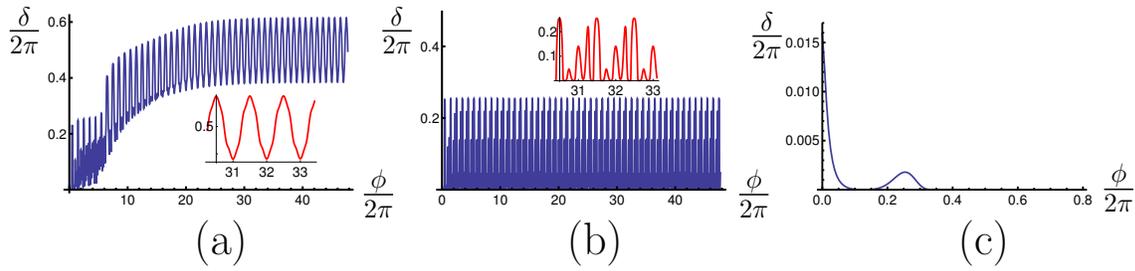
\begin{figure}
   \begin{center}
   \includegraphics[width=0.26\columnwidth]{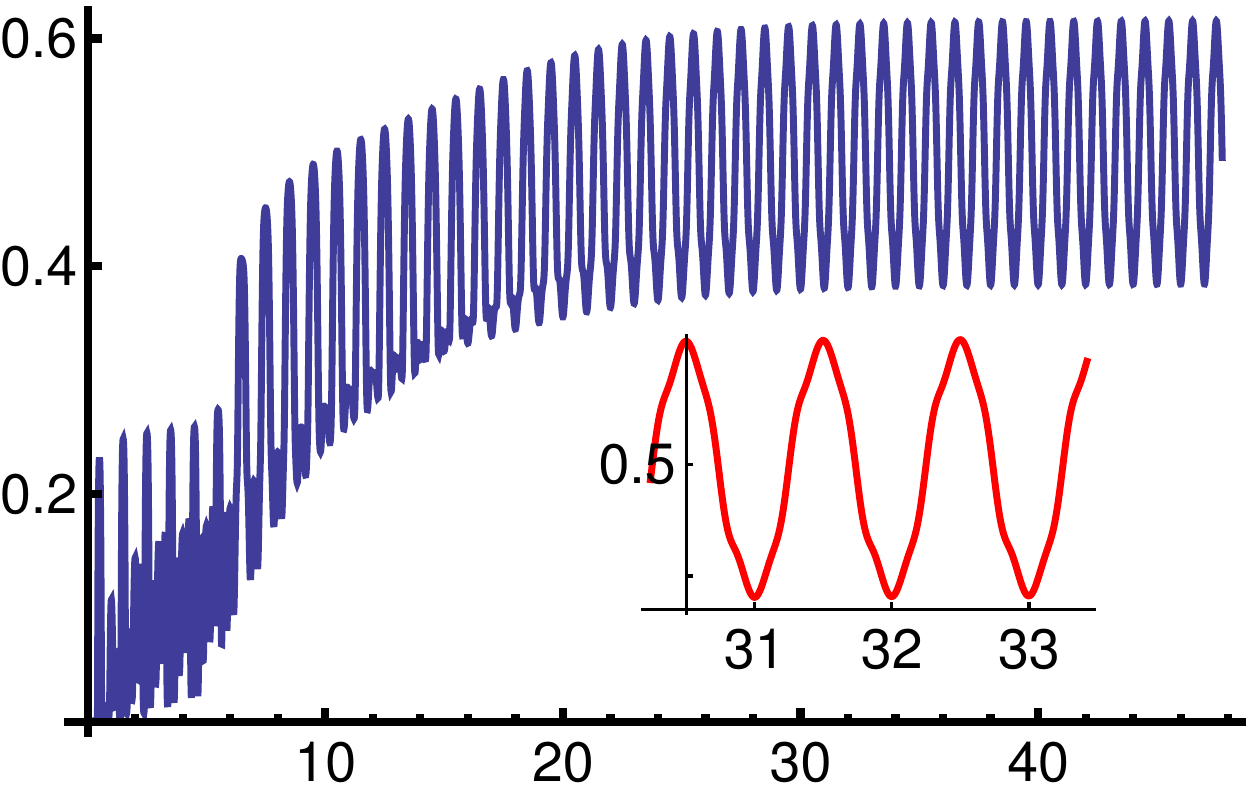}   
   \hspace{6mm}
   \includegraphics[width=0.26\columnwidth]{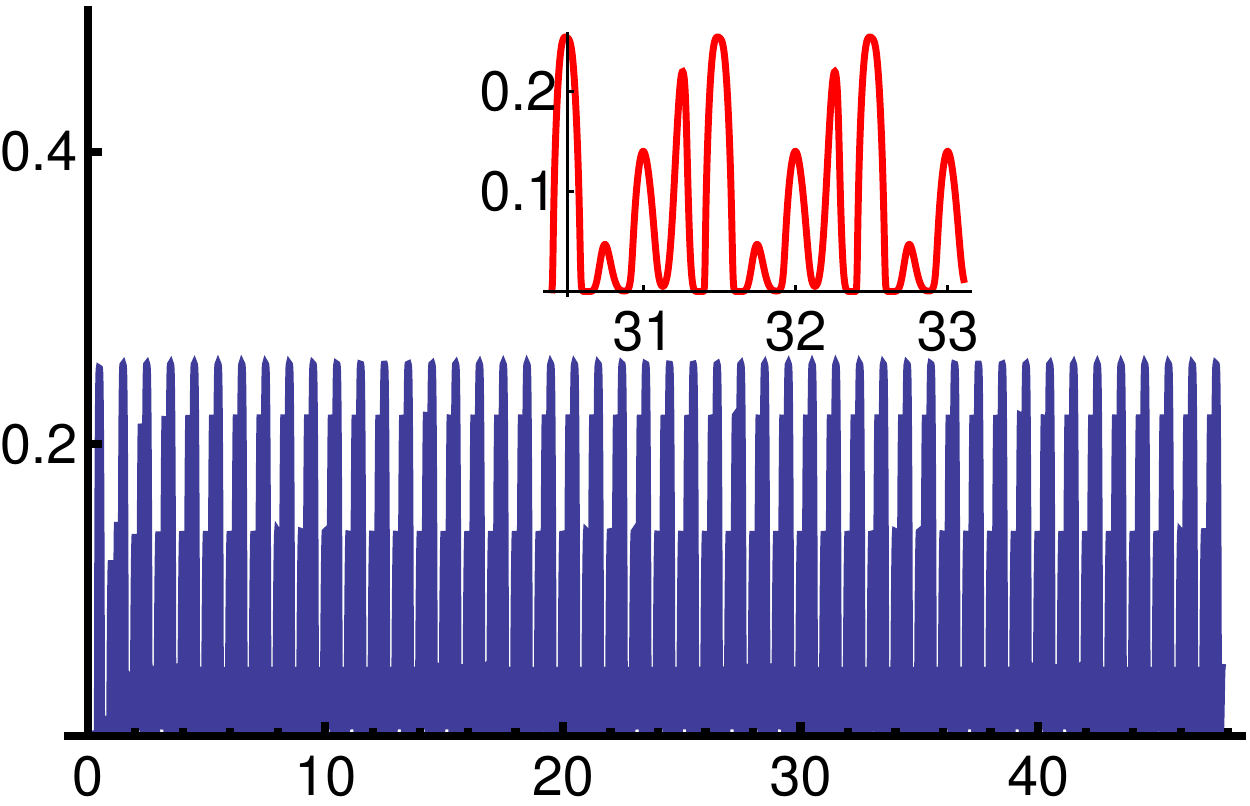}   
   \hspace{5.0mm}
   \includegraphics[width=0.26\columnwidth]{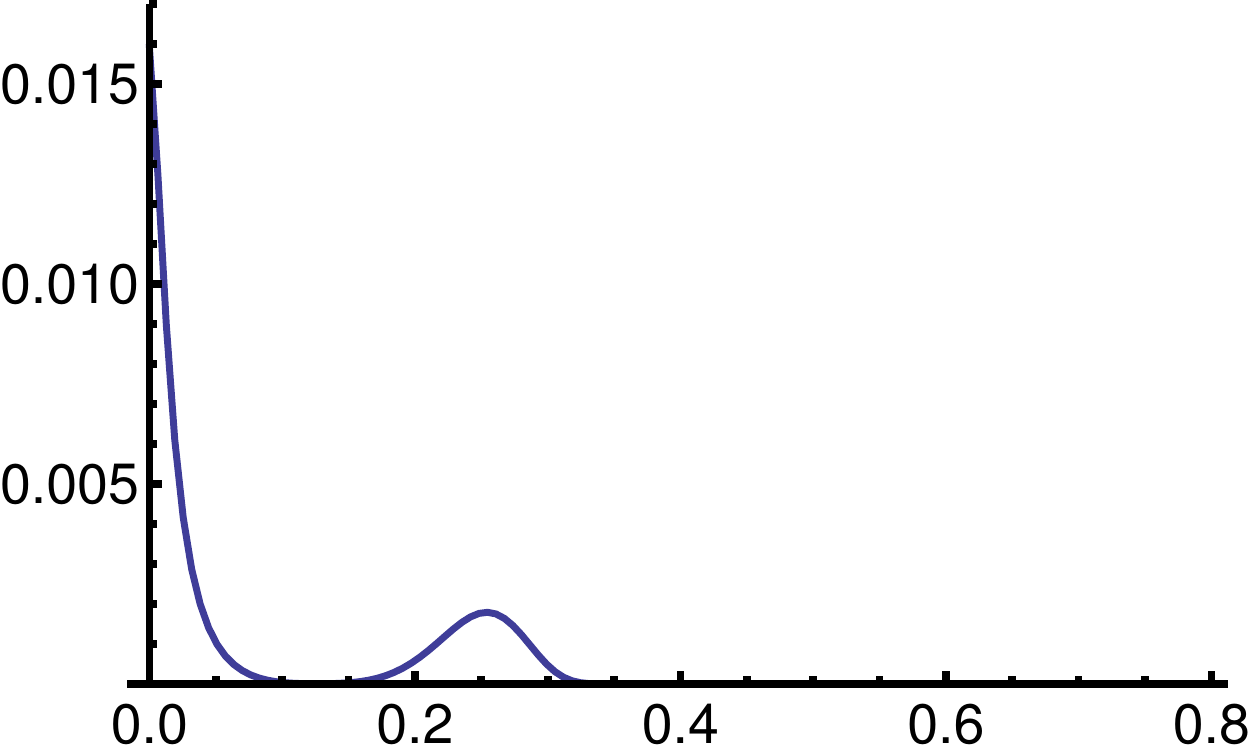}   
   \end{center}
   \begin{picture}(10,10)(0,0)
   \put(81,7){\Large{(a)}}
   \put(222,7){\Large{(b)}}
   \put(370,7){\Large{(c)}}

   \put(10,88){\large{$\frac{\delta}{2\pi}$}}
   \put(144,26){\large{$\frac{\phi}{2\pi}$}}

   \put(151,88){\large{$\frac{\delta}{2\pi}$}}
   \put(286,26){\large{$\frac{\phi}{2\pi}$}}

   \put(291,86){\large{$\frac{\delta}{2\pi}$}}
   \put(424,28){\large{$\frac{\phi}{2\pi}$}}

   \end{picture}
   \caption{Evolution of $\delta$ for the 4th harmonic with initial condition $\delta(\phi=0)=0.1$ and coefficient (a) $a^{(4)}=0.4$, (b) $a^{(4)}=0.8$, (c) $a^{(4)}=0.9$. (a), (b) $\delta$ evolves into different oscillating states. (c) $\delta$ evolves into the synchronized state. The insets show a small part of the plot in more detail.}
   \label{fig:harm4}
\end{figure}
Figure \ref{fig:harmstab} shows the stability of the synchronized state for the first 10 harmonics for a discrete range of coefficients when we choose the cosine term, i.e. $F_i^t(\phi_i)=F_0\big(1+a^{(n)} \cos{n \phi_i}\big)$.
\begin{figure}
   \begin{center}
   \includegraphics[width=0.55\columnwidth]{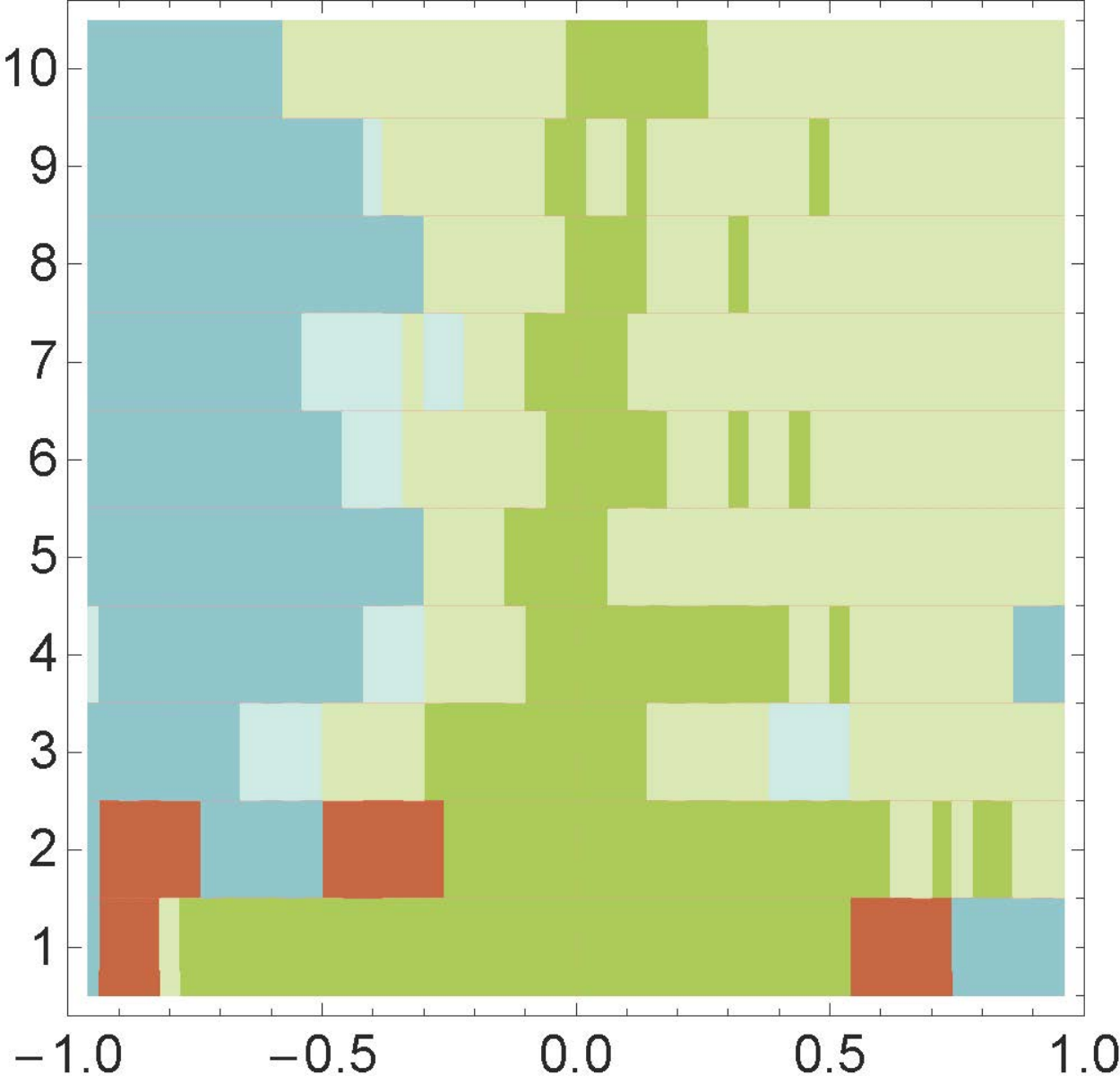} 
   \end{center}
   \begin{picture}(10,10)(0,0)
   \put(96,147){\large{$n$}}
   \put(225,12){\large{$a^{(n)}$}}
   \end{picture}
   \caption{Stabilities for range of coefficients and harmonics. Each horizontal band represents a harmonic, which is broken up into 49 blocks and each block represents a coefficient $a^{(n)}$. Orange blocks represent stability (i); dark blue blocks represent stability (ii); pale blue blocks represent stability (iii); pale green blocks represent stability (iv); and bright green blocks represent stability (v).} 
   \label{fig:harmstab}
\end{figure}
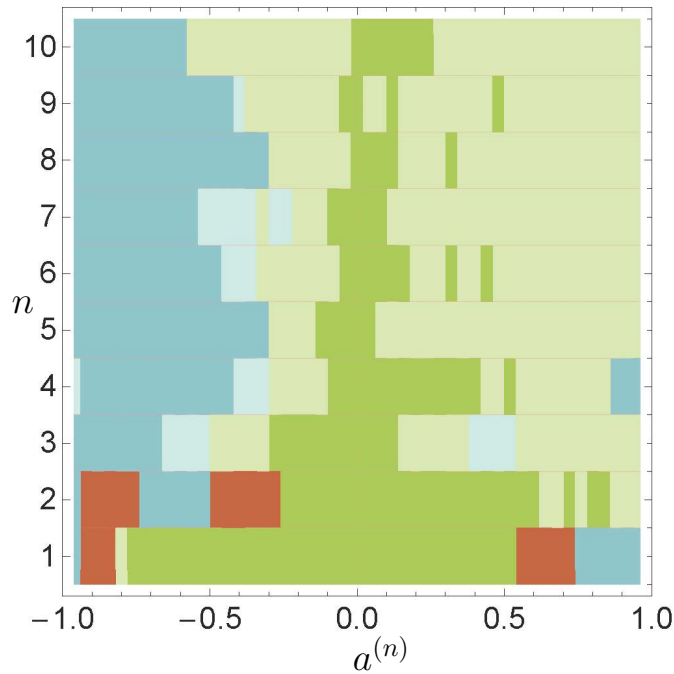
We see that there are more type (i) and (ii) force profiles for negative coefficients than for positive, showing that we are more likely to end up in the synchronized state when the coefficient is negative than when the coefficient is positive, for an arbitrary choice of harmonic. We see that for $n>2$ there is no type (i) behaviour, so we cannot guarantee that we will reach the synchronized states for these higher harmonics. The unstable type (v) band around $a^{(n)}=0$ gets narrower as $n$ increases, so we only need weak phase dependence to reach a synchronized state for higher harmonics, but the region of initial conditions which does lead to the synchronized state is very small. Even after reaching the synchronized state, it is likely that noise will drive $\delta$ away from the synchronized state. Usually the stability moves towards the lower types of stability (more stable types) as $|a^{(n)}|$ increases for each harmonic, although there are some exceptions. For example, when $n=1$ we can start in a type (i) region, then as $|a^{(1)}|$ increases we move into a type (ii) region. We also see a type (iii) region surrounded by type (iv) regions on both sides for $n=3$ and $a^{(3)}>0$. 

For each type of stability shown in figure \ref{fig:harmstab}, we select an arbitrary force profile and show the full initial condition phase diagram in figure \ref{fig:ICselection}. In the type (i) stable case, we see that oscillating states can still evolve (see also figure \ref{fig:ICphasediagram}), but the initial conditions for an oscillating state are not close to the synchronized case. In a few cases when we replace the cosine with sine, all initial conditions lead to the synchronized state, for example, the force profile $F^t(\phi) = F_0(1+b^{(1)}\sin{\phi})$ for $b^{(1)}\gtrsim0.6$, except for a very thin dotted white curve through the middle of the phase diagram indicating an unstable oscillating state. However, if we look at this oscillating state for long enough, after some time the numerical noise causes it to move into the synchronized state.

\begin{figure}
   \begin{center}
   \includegraphics[width=0.28\columnwidth]{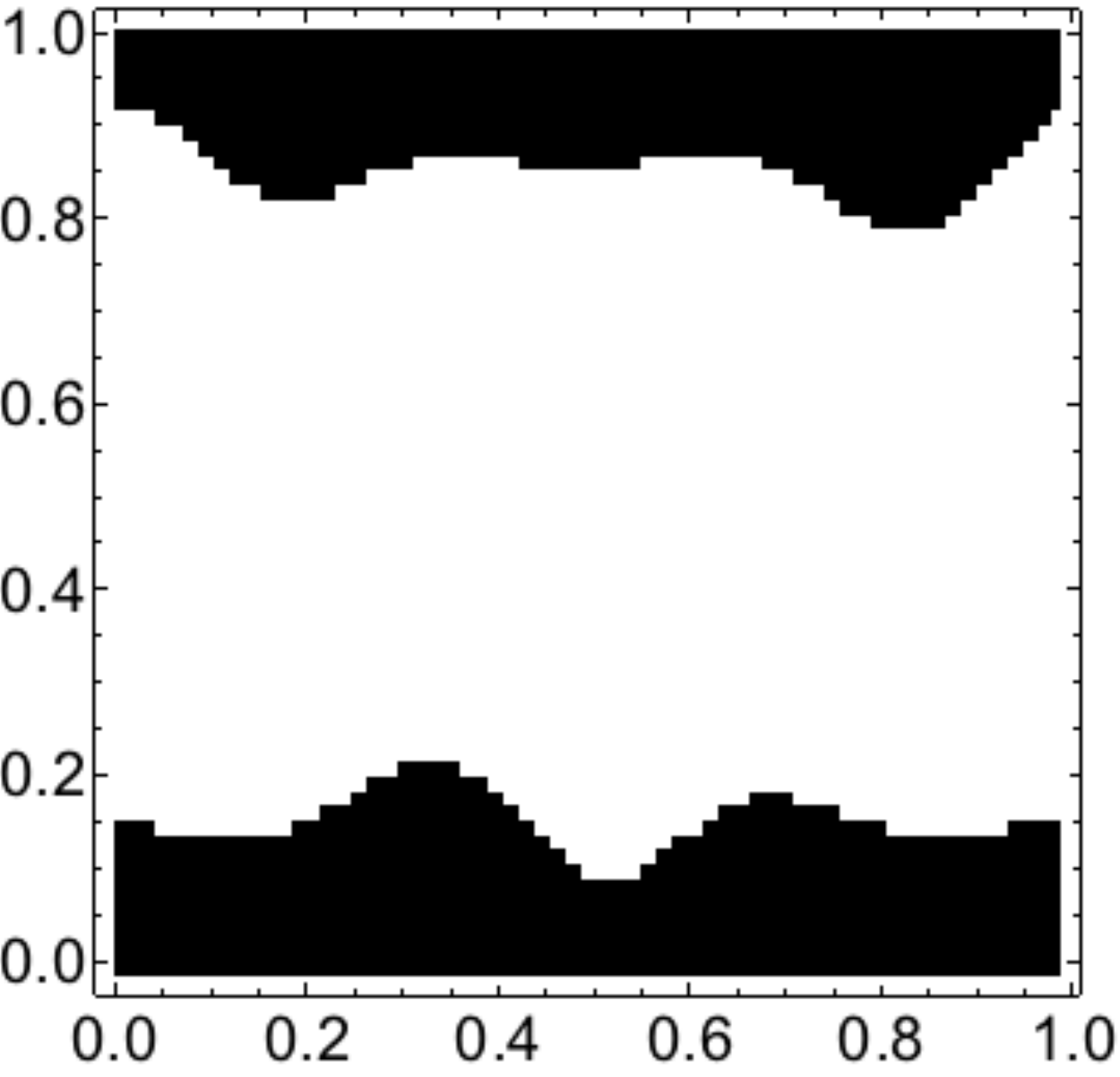}
   \hspace{5.0mm}
   \includegraphics[width=0.28\columnwidth]{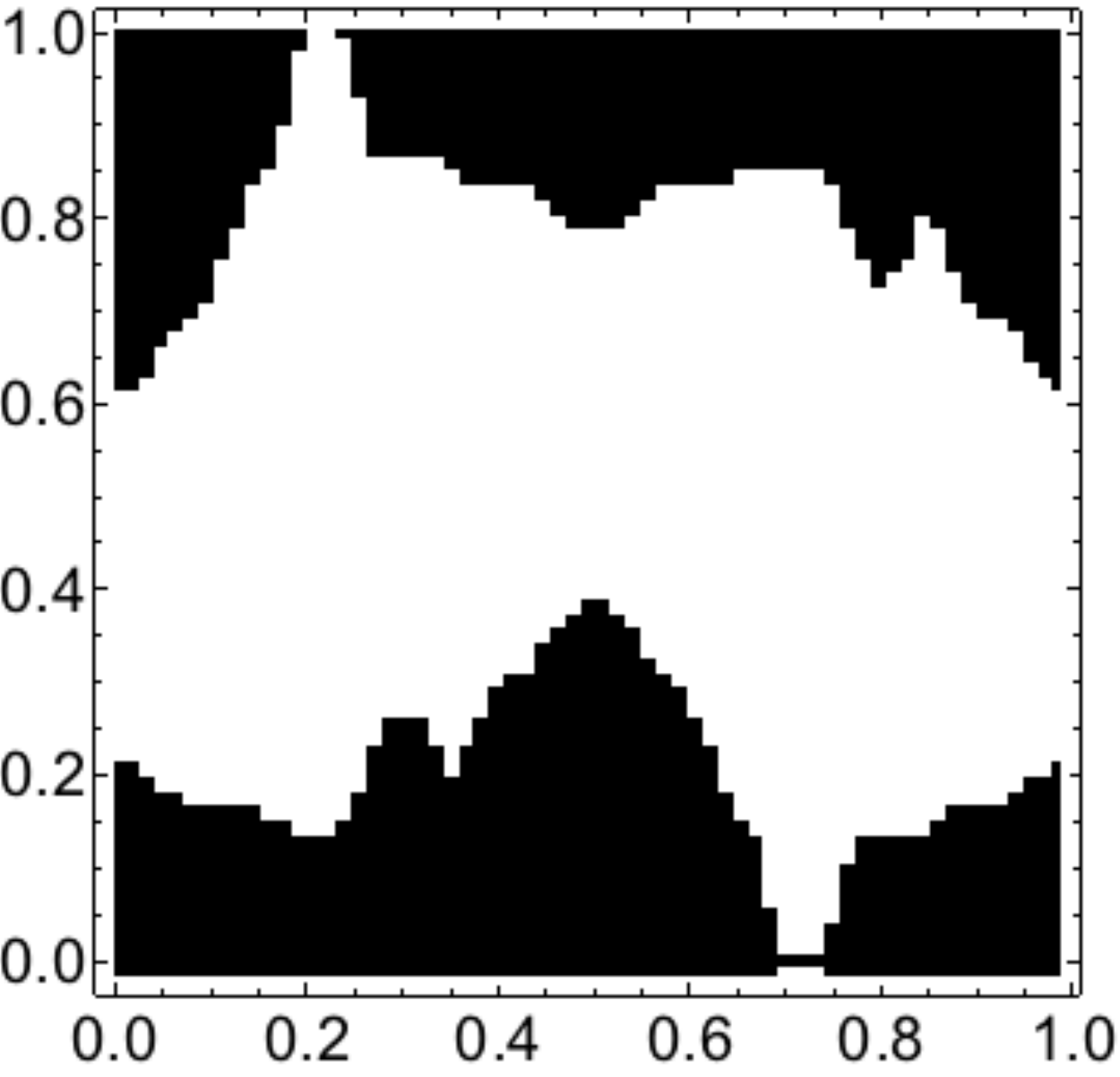}
   \hspace{5.0mm}
   \includegraphics[width=0.28\columnwidth]{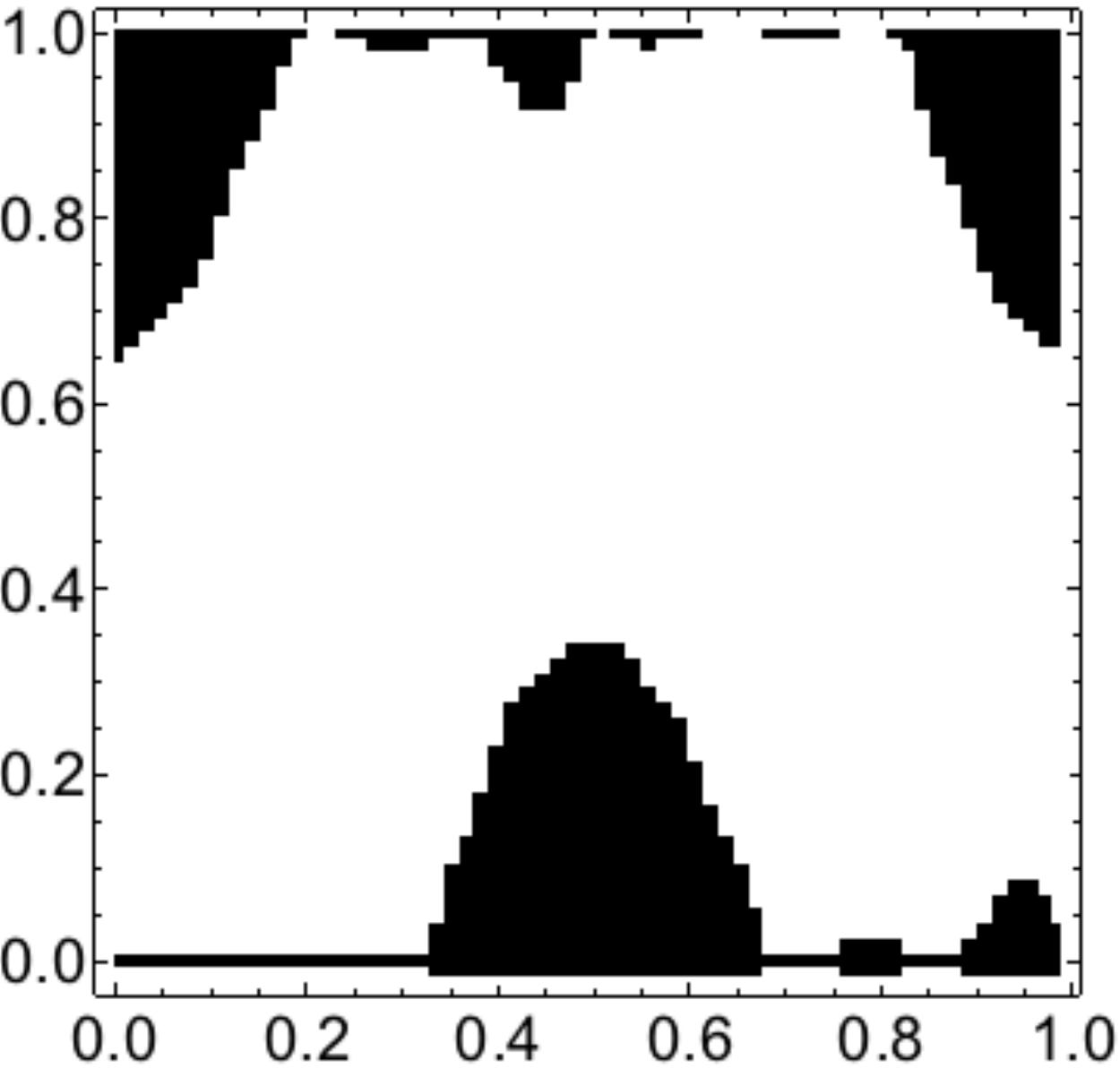} \\
   \vspace{12.0mm}
   \includegraphics[width=0.28\columnwidth]{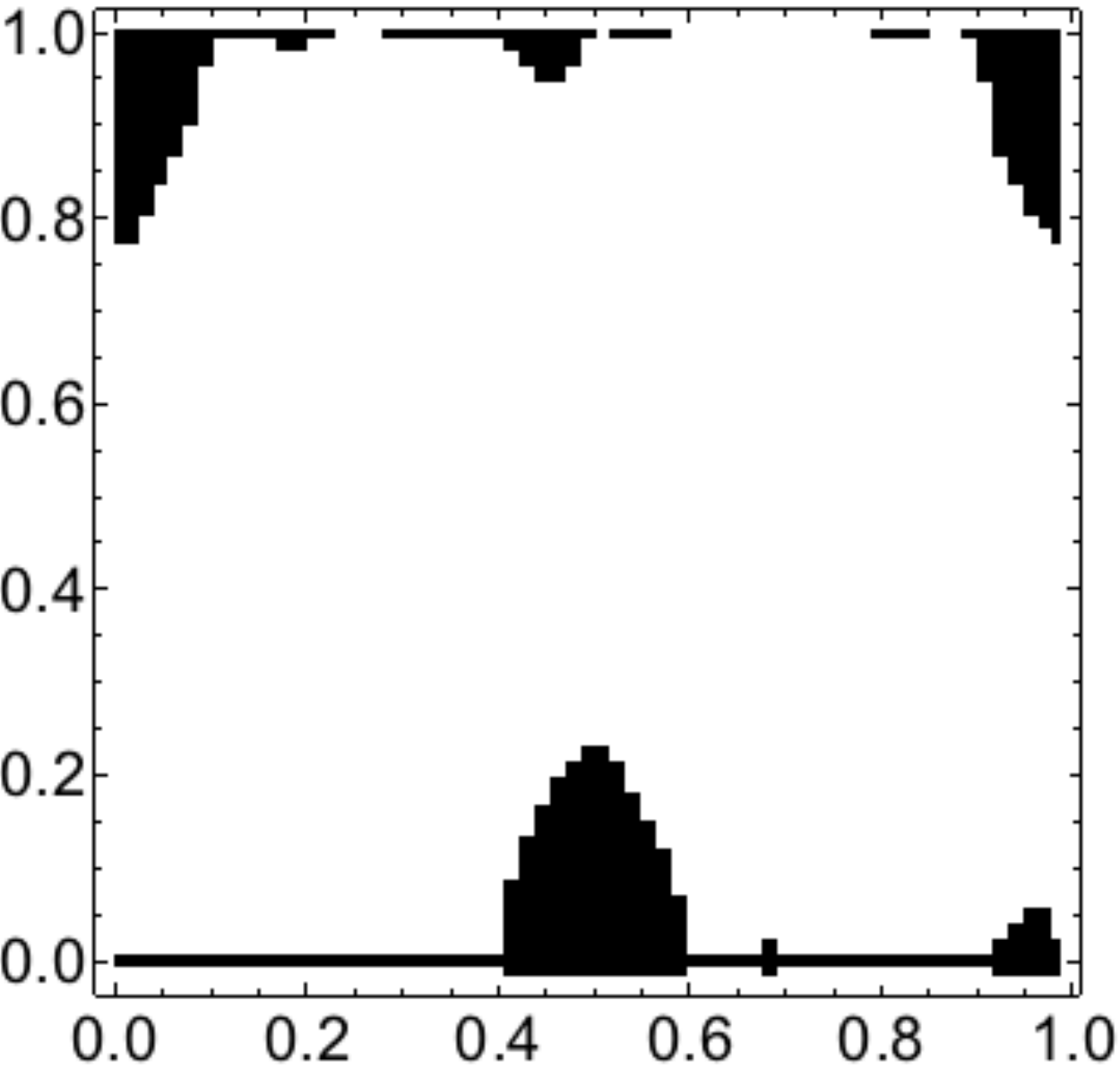}
   \hspace{7.0mm}
   \includegraphics[width=0.28\columnwidth]{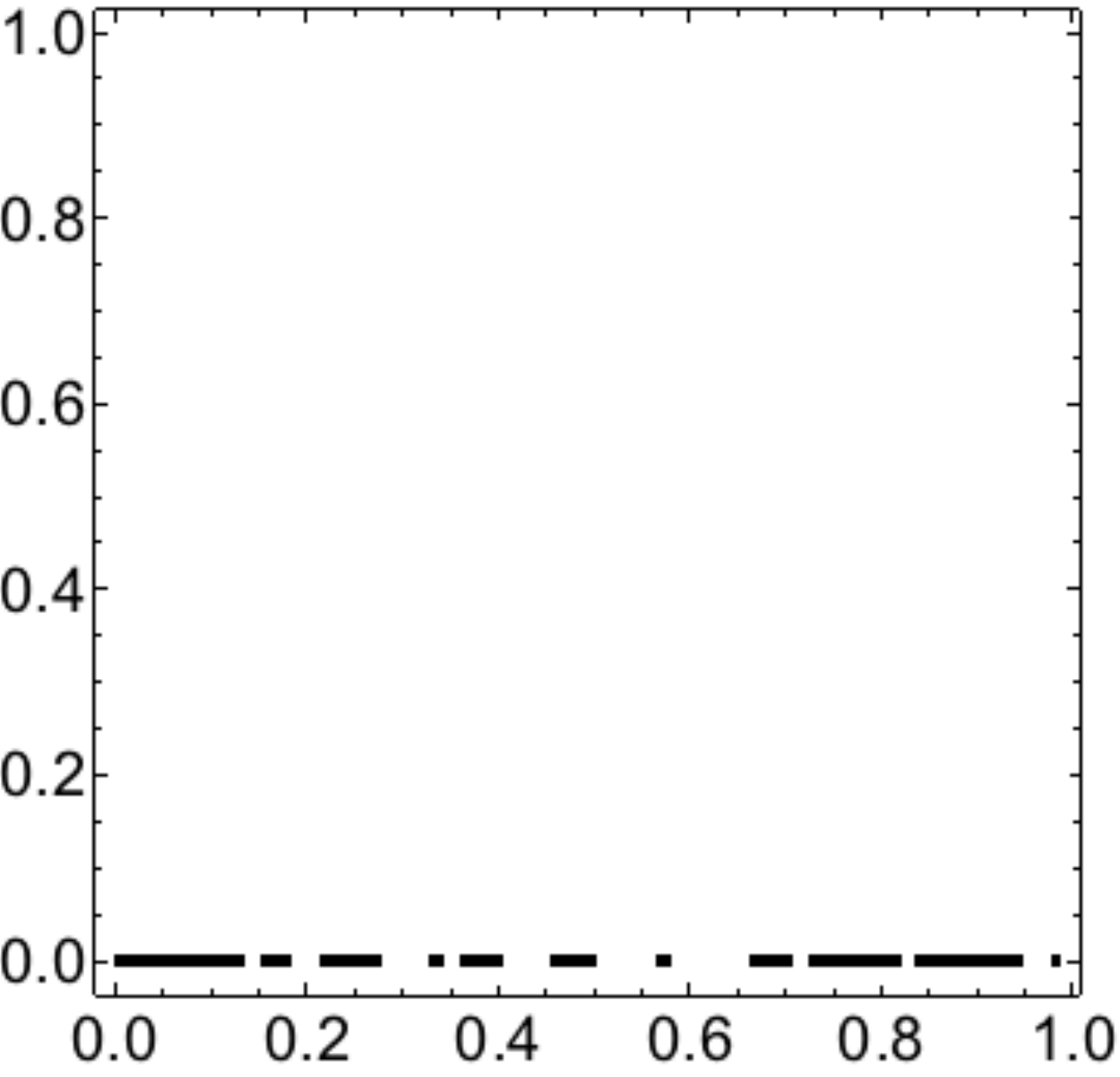}
   \end{center}
   \begin{picture}(10,10)(0,0)
   \put(73,161){\Large{(a)}}
   \put(219,161){\Large{(b)}}
   \put(367,161){\Large{(c)}}
   \put(143,2){\Large{(d)}}
   \put(297,2){\Large{(e)}}

   \put(0,288){\large{$\frac{\delta_0}{2\pi}$}}
   \put(122,169){\large{$\frac{\phi_0}{2\pi}$}}

   \put(148,288){\large{$\frac{\delta_0}{2\pi}$}}
   \put(269,169){\large{$\frac{\phi_0}{2\pi}$}}

   \put(295,288){\large{$\frac{\delta_0}{2\pi}$}}
   \put(416,169){\large{$\frac{\phi_0}{2\pi}$}}

   \put(71,134){\large{$\frac{\delta_0}{2\pi}$}}
   \put(192,12){\large{$\frac{\phi_0}{2\pi}$}}

   \put(225,134){\large{$\frac{\delta_0}{2\pi}$}}
   \put(345,12){\large{$\frac{\phi_0}{2\pi}$}}

   \end{picture}
   \caption{Initial condition phase diagrams for each type of stability and force profile $F^t(\phi) = F_0(1+a^{(n)}\cos{n\phi})$. (a) Type (i) stable diagrams for $n = 2$ and $a^{(n)} = -0.44$. (b) Type (ii) stability for $n = 7$ and $a^{(n)} = -0.8$. (c) Type (iii) stability for $n = 7$ and $a^{(n)} =-0.44$. (d) Type (iv) stability for $n = 8$ and $a^{(n)} =-0.24$. (e) Type (v) stability for $n = 6$ and $a^{(n)} = -0.04$.}
   \label{fig:ICselection}
\end{figure}
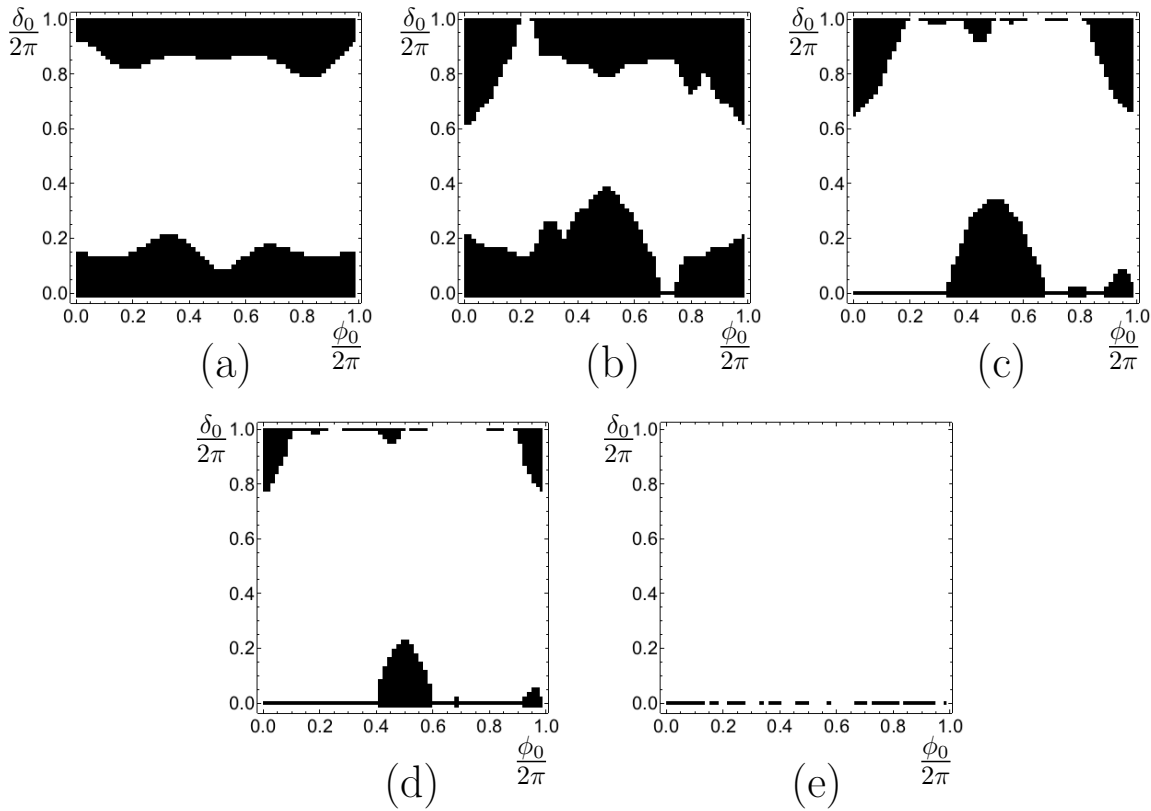

We see from the bright green band down the center of figure \ref{fig:harmstab} that we need some form of phase dependence in order to achieve synchronization. If we choose a constant driving force then $\delta$ evolves into an oscillating state for all initial conditions. The synchronized state is unstable, so even if we start in the synchronized state, a small amount of numerical noise can drive the system into the oscillating state.
Friedrich {\it et al.} showed that synchronization can occur with constant forcing when the direction of rotation of the beads is reversed (equivalent to $H \to -H$ and reversing swimming direction) but here we focus on the case $F_i^t>0$ and $H>0$.

The inspiration for our run-and-tumble model in reference \cite{bg:2012} came from the initial condition phase diagrams. To see run-and-tumble we want to start in a stable synchronized state, then allow noise to move us temporarily into a white region of the phase diagram, before moving back into a black region. In reference \cite{bg:2012}, we allowed noise to vary the coefficients $a_i^{(1)}$, so that a black square in the noiseless  phase diagram can change to a white square when the instantaneous effect of the noise changes the value of  $a_i^{(1)}$, then changes back to a black square when the instantaneous effect of the noise is smaller. This allows us to start in the synchronized state (with fluctuations due to the noise), then move away from the synchronized state when the instantaneous noise is large and we are at a suitable point in the phase diagram, then move back into the fluctuating synchronized state when the instantaneous noise is small. In reference \cite{bg:2012}, we chose to work with the first harmonic and use $a^{(1)}=0.7+\zeta$, where $\zeta$ is the noise term. We chose the value 0.7 because it is in the stability type (i) region, but lies close to the type (ii) region. In the fluctuating synchronized state, when we move through the values of $\phi$ which are surrounded by white squares in the type (ii) phase diagram, there is the possibility to move into an oscillating state, but there are still plenty of black squares surrounding the line $\delta=0$, so we can have long periods in the {\em run} phase.

The noise causes fluctuations of the position in the phase diagram, and this could also be a cause of run-and-tumble behaviour. For example, consider the phase diagram in figure \ref{fig:ICselection}(a). If we are in the synchronized state and noise is small enough such that fluctuations in $\delta$ are within the black region, then tumbles will not occur. If the noise is larger, so $\delta$ fluctuations move into the white region, then the cell could begin to move into the oscillating state, and after a few oscillations noise could kick the oscillations into the black region and the cell would move back towards a synchronized state. Elsewhere, we will consider the effects of adding noise to $\dot{\phi}$ and $\dot{\delta}$, without any noise in the coefficient, to see if we obtain run-and-tumble behaviour this way and compare the statistics to the run-and-tumble obtained when noise is added to the coefficient.

\subsection{Mismatched coefficients}
If we start in the synchronized state, then
$$\left. \frac{{\rm d}\delta}{{\rm d}\phi}\right|_{\delta=0} = \Bigg(\frac{F_l^t-F_r^t}{F_l^t+F_r^t}\Bigg) \Phi({\phi}),$$
which is non-zero for $F_l^t \not= F_r^t$, so the system does not stay in the synchronized state.
We focus on the first harmonic and consider the case $F_l^t = F_0(1+a_l \cos\phi_l)$ and $F_r^t = F_0(1+a_r \cos\phi_r)$ where $a_l \not= a_r$ (and we have dropped the upper index on the coefficient). When $a_l=-a_r$ and $|a_i|\gtrsim0.6$, synchronization is frustrated and $\delta$ oscillates about $2 \pi n$, $n \in \mathbb{Z}$, shown in figure \ref{fig:MC}(a) for $-a_l=a_r=0.7$. The orientation of the cell drifts, shown in figure \ref{fig:MC}(b) so the cell swims along a curved trajectory. For $|a_i|<0.6$, the centre of $\delta$ oscillations drifts away from the synchronized state, but remains close to $2\pi n$. Swapping the signs of the coefficients swaps the direction of the orientation drift.

When the coefficients have the same sign but different magnitudes, $\delta$ oscillates about $(2n +1)\pi$ and the orientation oscillates about some fixed value. Figure \ref{fig:MC}(c) shows the evolution of $\delta$ for $a_l=0.7, a_r=0.8$. This type of behaviour is also seen for some choices of equal coefficients, for example, in figure \ref{fig:devolutionS1}(e), (f) where $b_l^{(1)}=b_r^{(1)}=-0.1$. Choices of coefficients which give this type of behaviour can be used to model a mutant of {\it Chlamydomonas} which swims with antiphase synchrony \cite{Goldstein2}. When the model swims with antiphase beating, the beat frequency is higher than when it swims with in phase beating, which has been observed in real {\it Chlamydomonas} cells \cite{Goldstein2}.

\begin{figure}
   \begin{center}
   \hspace{1.0mm}
   \includegraphics[width=0.27\columnwidth]{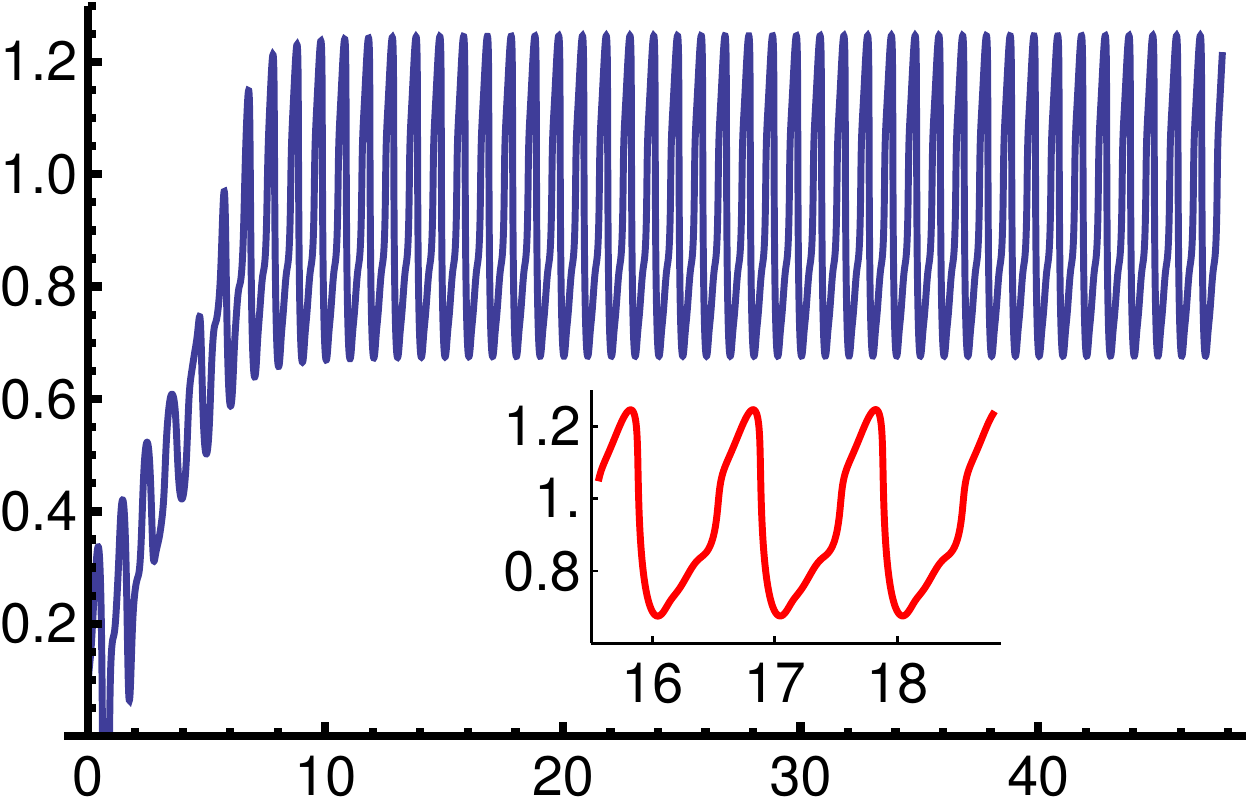}
   \hspace{4.2mm}
   \includegraphics[width=0.27\columnwidth]{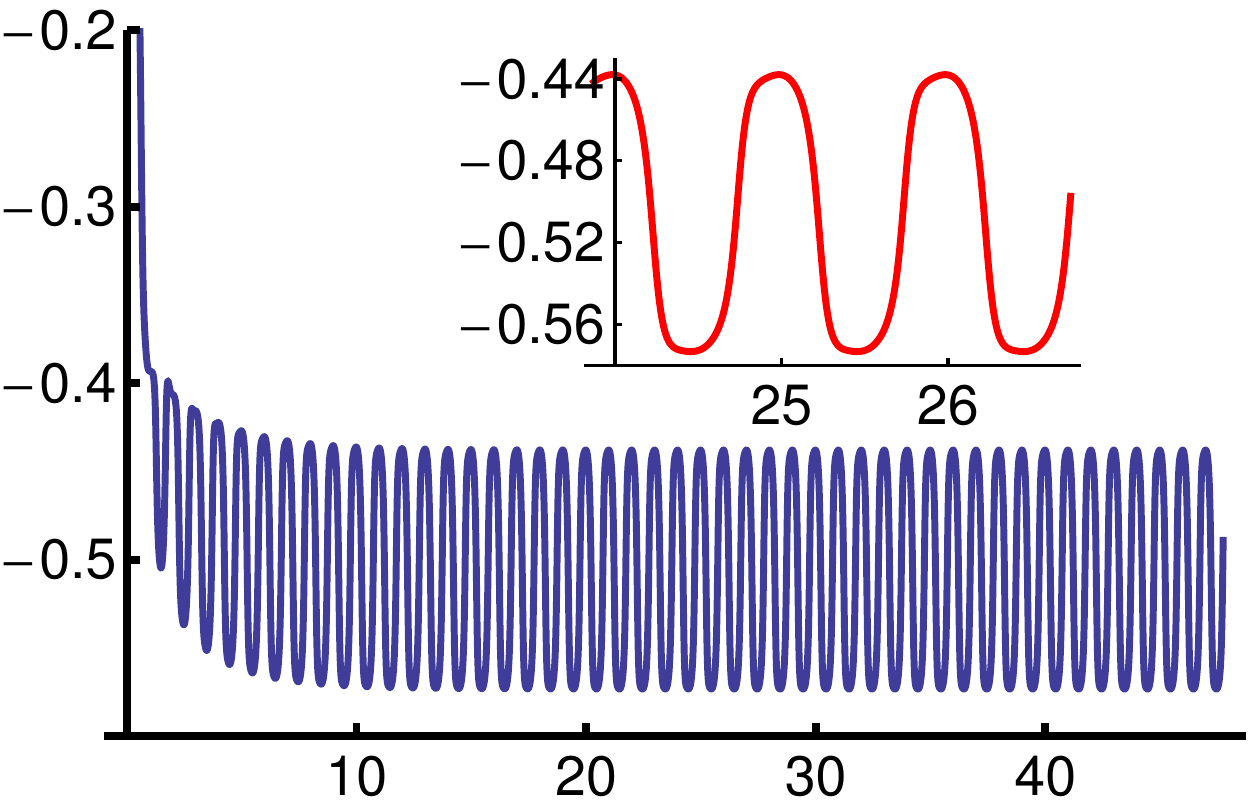}
   \hspace{4.2mm}
   \includegraphics[width=0.27\columnwidth]{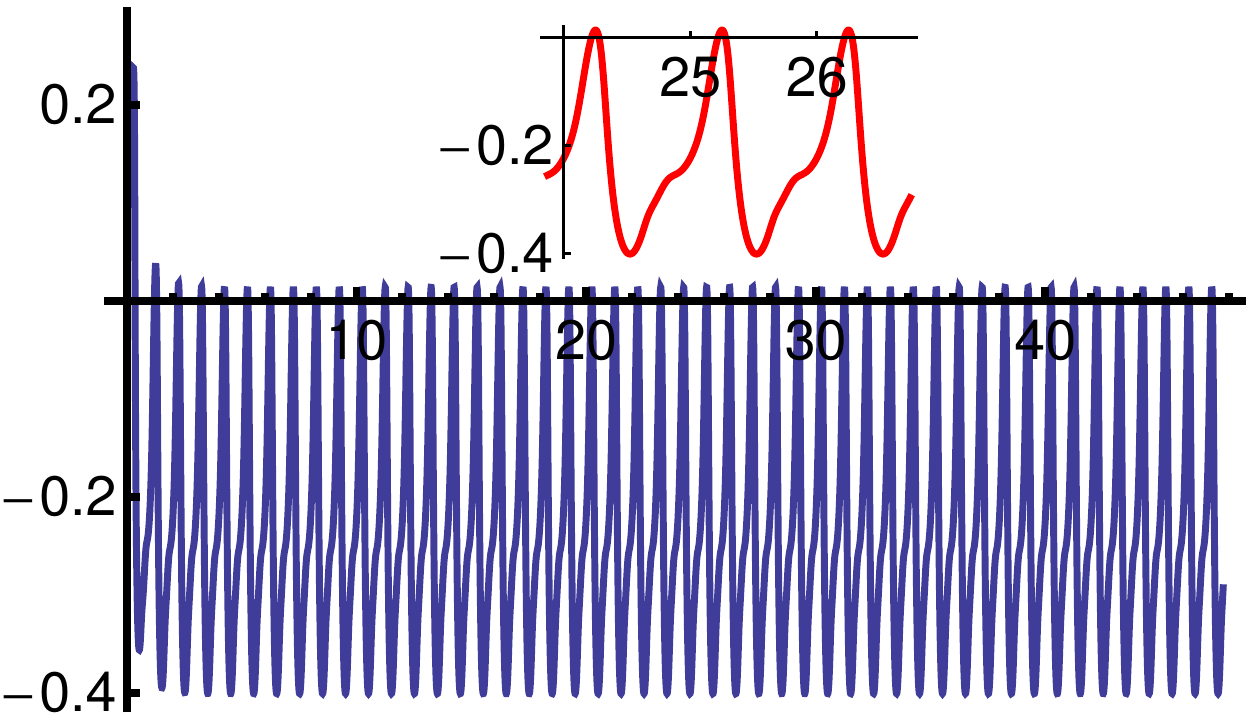} \\
   \vspace{8.0mm}
   \hspace{0.05mm}
   \includegraphics[width=0.278\columnwidth]{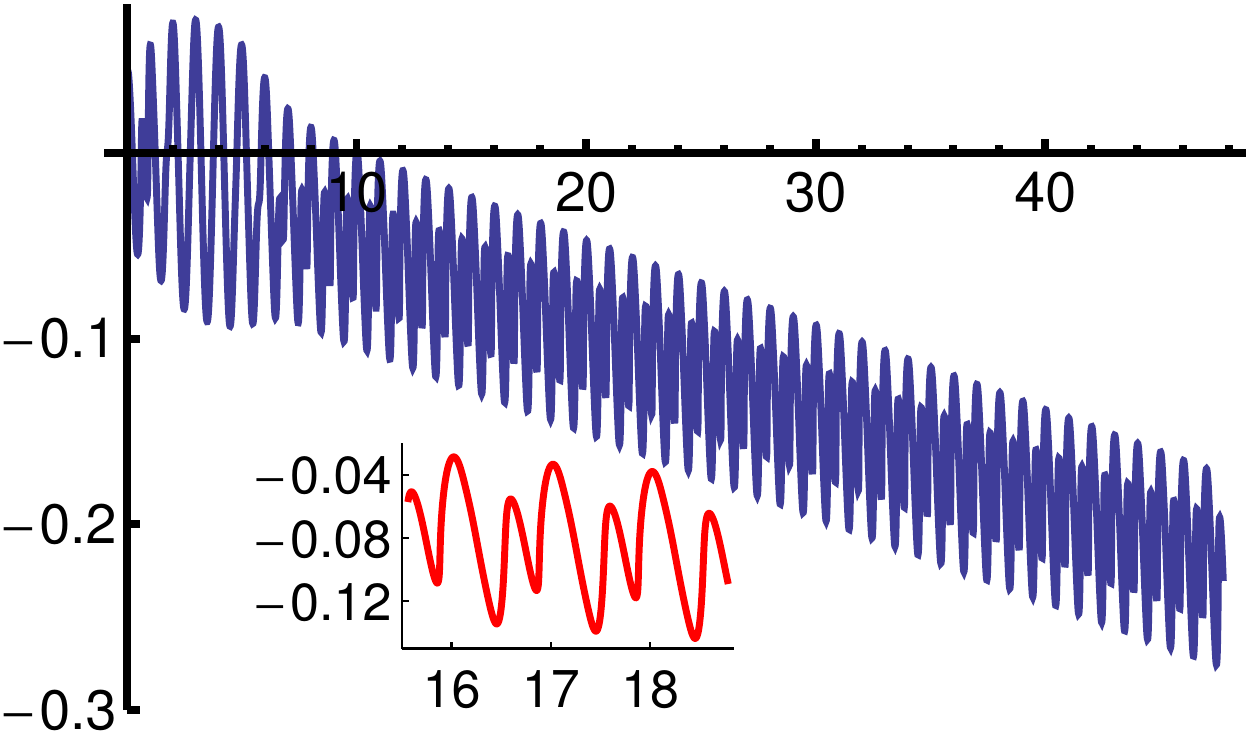}
   \hspace{3.0mm}
   \includegraphics[width=0.28\columnwidth]{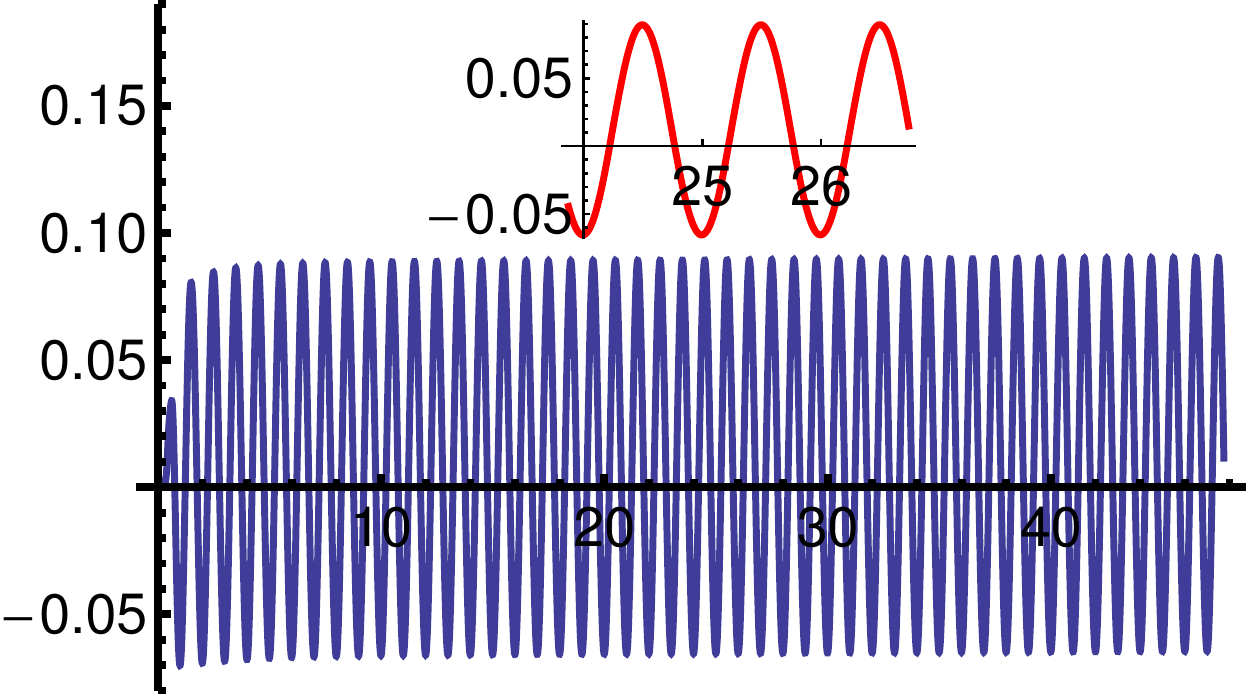}
   \hspace{4.2mm}
   \includegraphics[width=0.27\columnwidth]{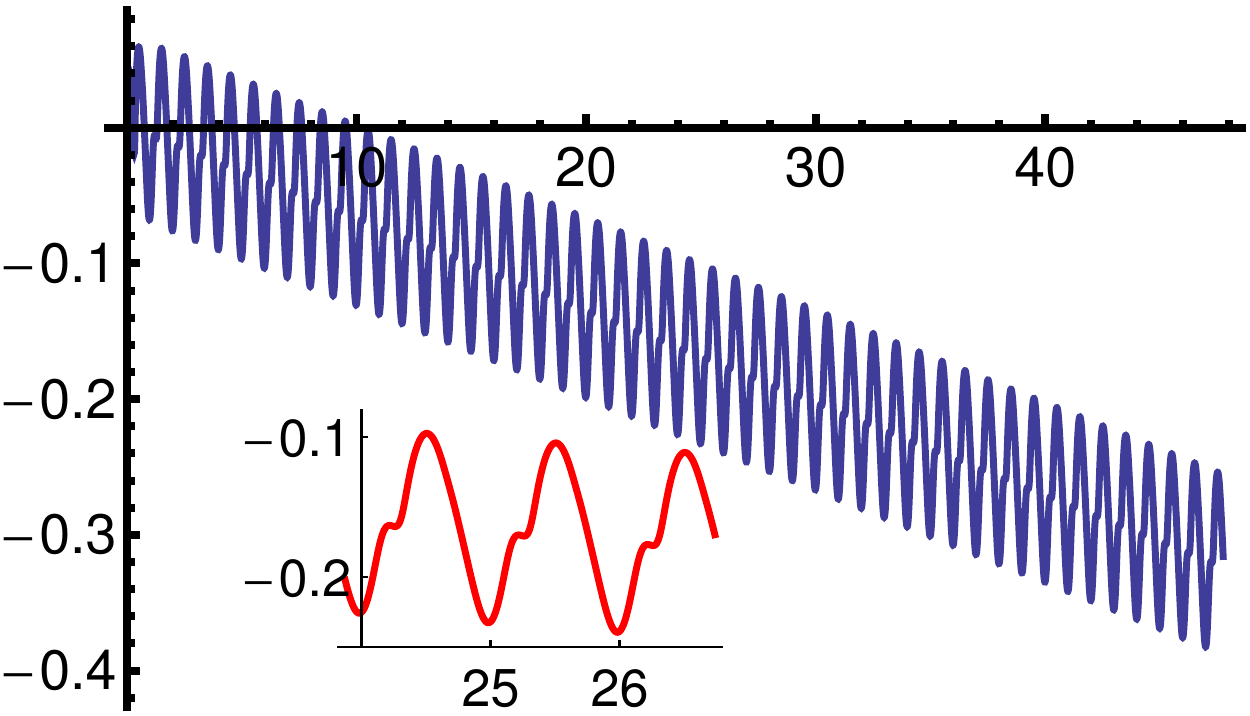}
   \end{center}
   \begin{picture}(10,10)(0,0)
   \put(83,107){\Large{(a)}}
   \put(83,9){\Large{(b)}}
   \put(226,107){\Large{(c)}}
   \put(226,9){\Large{(d)}}
   \put(368,107){\Large{(e)}}
   \put(368,9){\Large{(f)}}

   \put(12,188){\large{$\frac{\delta}{2\pi}$}}
   \put(148,125){\large{$\frac{\phi}{2\pi}$}}

   \put(16,84){\large{$\frac{\theta}{2\pi}$}}
   \put(149,80){\large{$\frac{\phi}{2\pi}$}}

   \put(155,188){\large{$\frac{\delta}{2\pi}$}}
   \put(288,125){\large{$\frac{\phi}{2\pi}$}}

   \put(165,94){\large{$\frac{\theta}{2\pi}$}}
   \put(290,41){\large{$\frac{\phi}{2\pi}$}}

    \put(295,182){\large{$\frac{\delta}{2\pi}$}}
    \put(430,159){\large{$\frac{\phi}{2\pi}$}}

   \put(304,85){\large{$\frac{\theta}{2\pi}$}}
   \put(431,78){\large{$\frac{\phi}{2\pi}$}}

   \end{picture}

   \caption{(a), (c), (e) Evolution of $\delta$ and (b), (d), (f) corresponding orientation $\theta$ for force profiles $F_{i}^t(\phi_i)=F_0(1+a_i \cos{\phi_i})$ and initial conditions $\delta(\phi=0)=0$, $\theta(\phi=0)=0$ and with (a), (b) $-a_l=a_r=0.7$; (c), (d) $a_l=0.7, a_r=0.8$; (e), (f) $a_l=-0.7, a_r=0.8$. Each inset shows a small part of the main plot in more detail.}
   \label{fig:MC}
\end{figure}
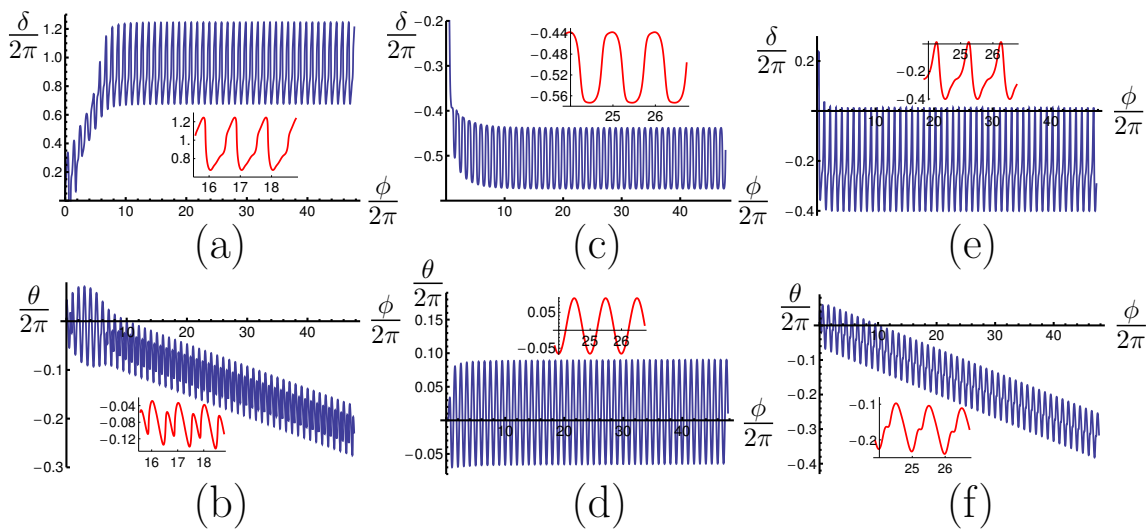

When the coefficients have opposite signs and different magnitudes, there are two main types of behaviour that occur. For $|a_l|<|a_r|$ and ${\rm sgn}(a_l)=-1$ or $|a_r|<|a_l|$ and ${\rm sgn}(a_r)=-1$, then $\delta$ oscillates periodically and the corresponding orientation drifts in the negative direction in the former case and in the positive direction in the latter case. This is shown in figure \ref{fig:MC}(e), (f) for $a_l=-0.7, a_r=0.8$. For $|a_l|<|a_r|$ and ${\rm sgn}(a_l)=+1$ or $|a_r|<|a_l|$ and ${\rm sgn}(a_r)=+1$, then oscillations in phase difference, $\delta$, drift in the negative (positive) direction and the orientation drifts in the positive (negative) direction in the former (latter) case. Figure \ref{fig:deldrift} shows examples of this this case where one bead completes more cycles than the other bead. There is a difference in the mean angular velocity of the two beads without choosing a different $F_0$ for each bead.

\begin{figure}
   \begin{center}
   \hspace{0.5mm}
   \includegraphics[width=0.35\columnwidth]{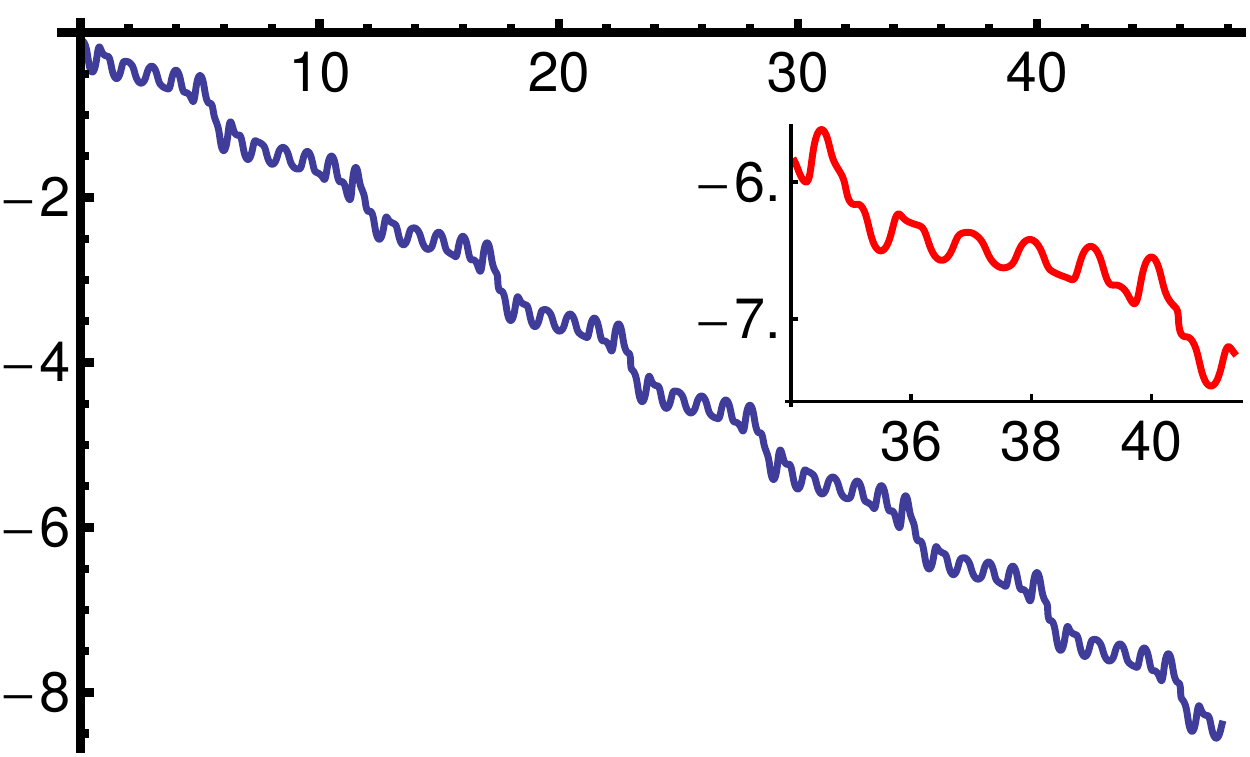} 
   \hspace{14.0mm}
   \includegraphics[width=0.34\columnwidth]{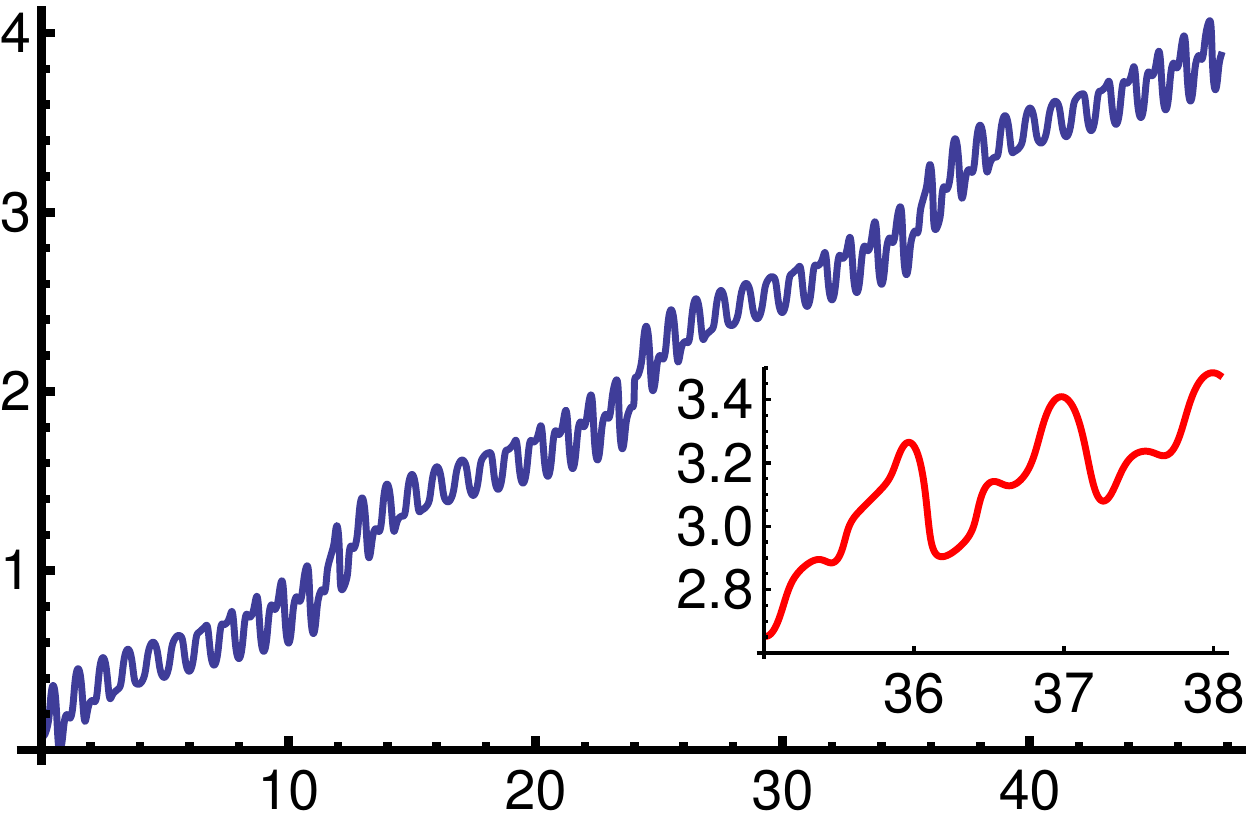}  \\ 
   \vspace{10.0mm}
   \includegraphics[width=0.37\columnwidth]{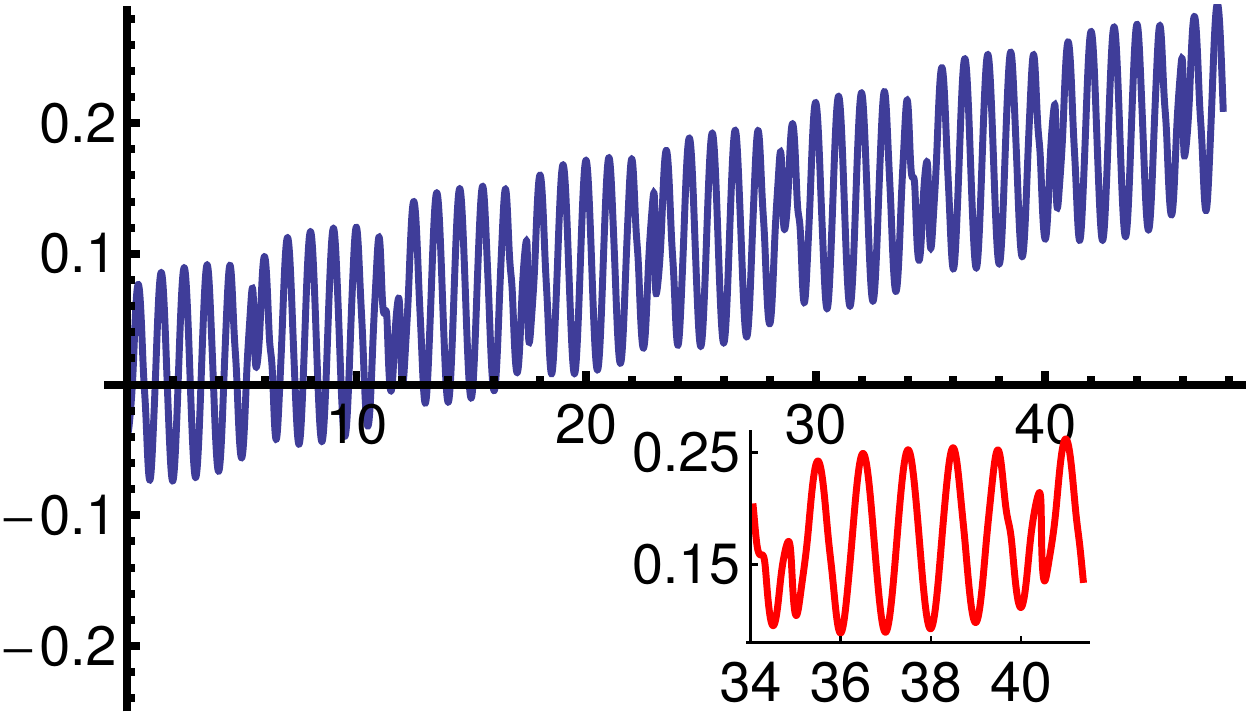}
   \hspace{9.0mm}
   \includegraphics[width=0.37\columnwidth]{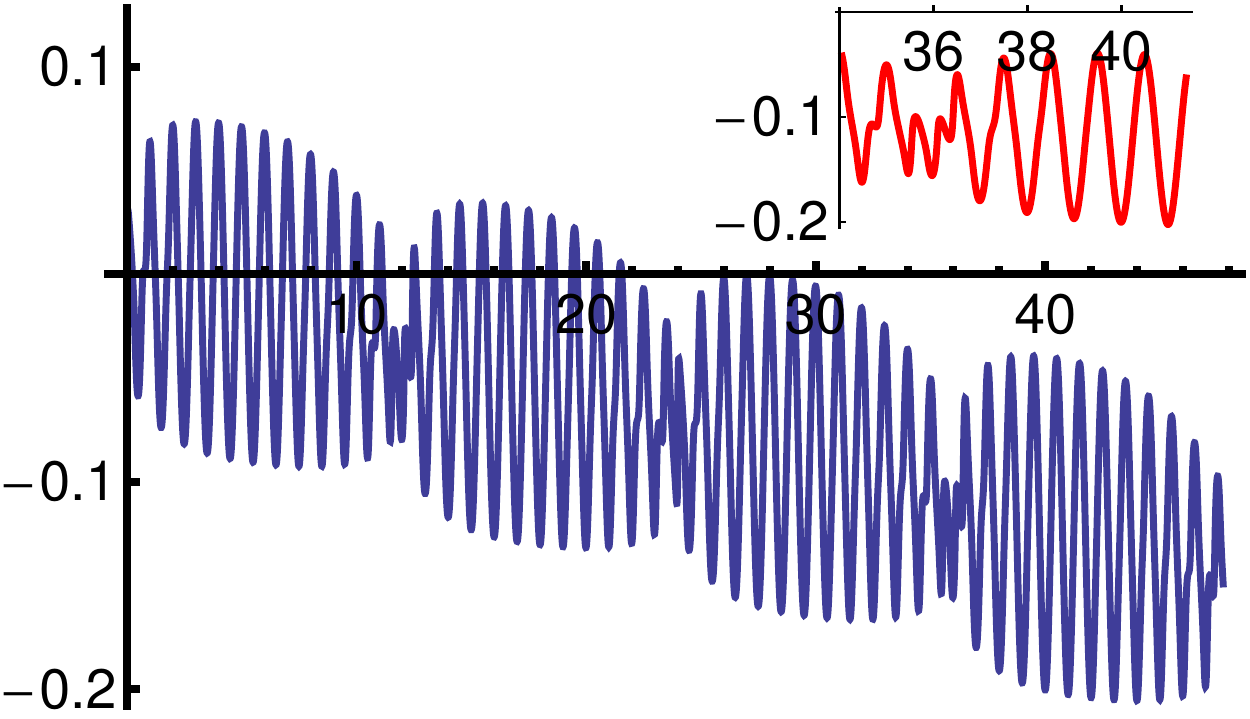}
   \end{center}

   \begin{picture}(10,10)(0,0)
   \put(128,136){\Large{(a)}}
   \put(128,6){\Large{(b)}}
   \put(325,136){\Large{(c)}}
   \put(325,6){\Large{(d)}}

   \put(40,234){\large{$\frac{\delta}{2\pi}$}}
   \put(207,236){\large{$\frac{\phi}{2\pi}$}}

   \put(32,108){\large{$\frac{\theta}{2\pi}$}}
   \put(209,65){\large{$\frac{\phi}{2\pi}$}}

   \put(239,239){\large{$\frac{\delta}{2\pi}$}}
   \put(406,154){\large{$\frac{\phi}{2\pi}$}}

   \put(228,109){\large{$\frac{\theta}{2\pi}$}}
   \put(408,80){\large{$\frac{\phi}{2\pi}$}}

   \end{picture}

   \caption{Drifting $\delta$ and corresponding orientation for (a), (b) $F_l^t=F_0(1+0.2\cos{\phi_l})$ and $F_r^t=F_0(1-0.6\cos{\phi_r})$; (c), (d) $F_l^t=F_0(1-0.4\cos{\phi_l})$ and $F_r^t=F_0(1+0.3\cos{\phi_r})$. The insets show a small part of the plot in more detail.}
   \label{fig:deldrift}
\end{figure}
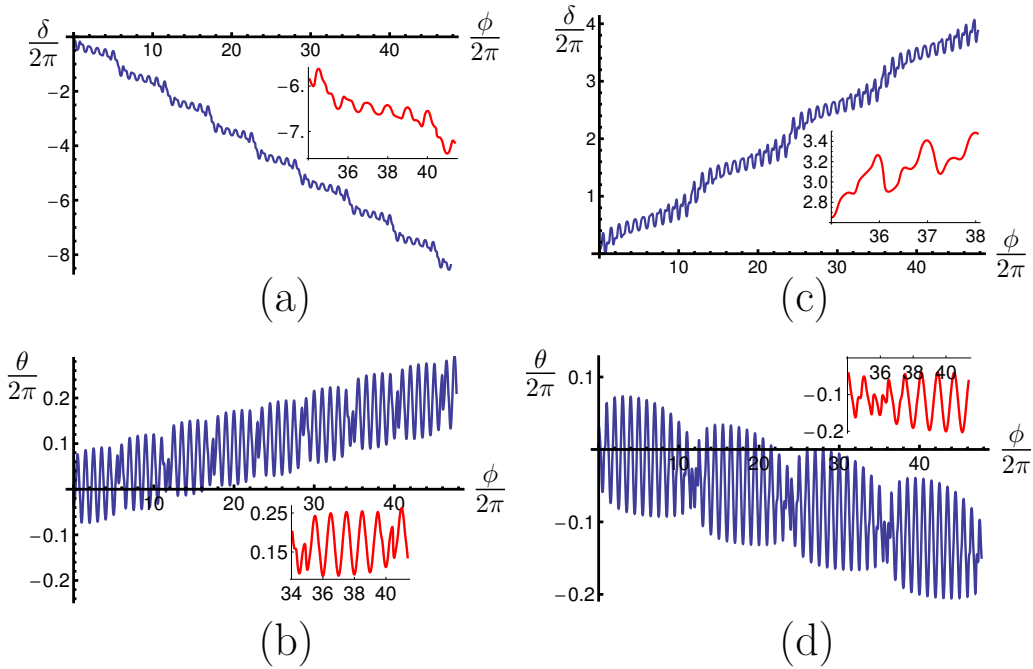

\subsection{Combinations of harmonics}
So far we have considered force profiles with only one harmonic term. Now we consider driving forces with contributions from two harmonics. For simplicity we choose equal profiles for the left and right beads, $F_l^t(\phi) = F_r^t(\phi)=F^t(\phi)$, and of the form $F^t(\phi) = F_0(1+a^{(m)}\cos{m\phi}+a^{(n)}\cos{n\phi})$, $m \not= n$ and with the $a^{(i)}$'s chosen such that $F^t(\phi)>0$ for all real $\phi$. Figure \ref{fig:harm12} shows the stability of the synchronized state for $m=1, n=2$ and $m=1, n=3$. Each grid square represents a choice of driving force with coefficients $(a^{(1)},a^{(j)})$, $j = 2,3$ and the colour represents the stability of the synchronized state for that particular driving force.
\begin{figure}
   \begin{center}
   \includegraphics[width=0.36\columnwidth]{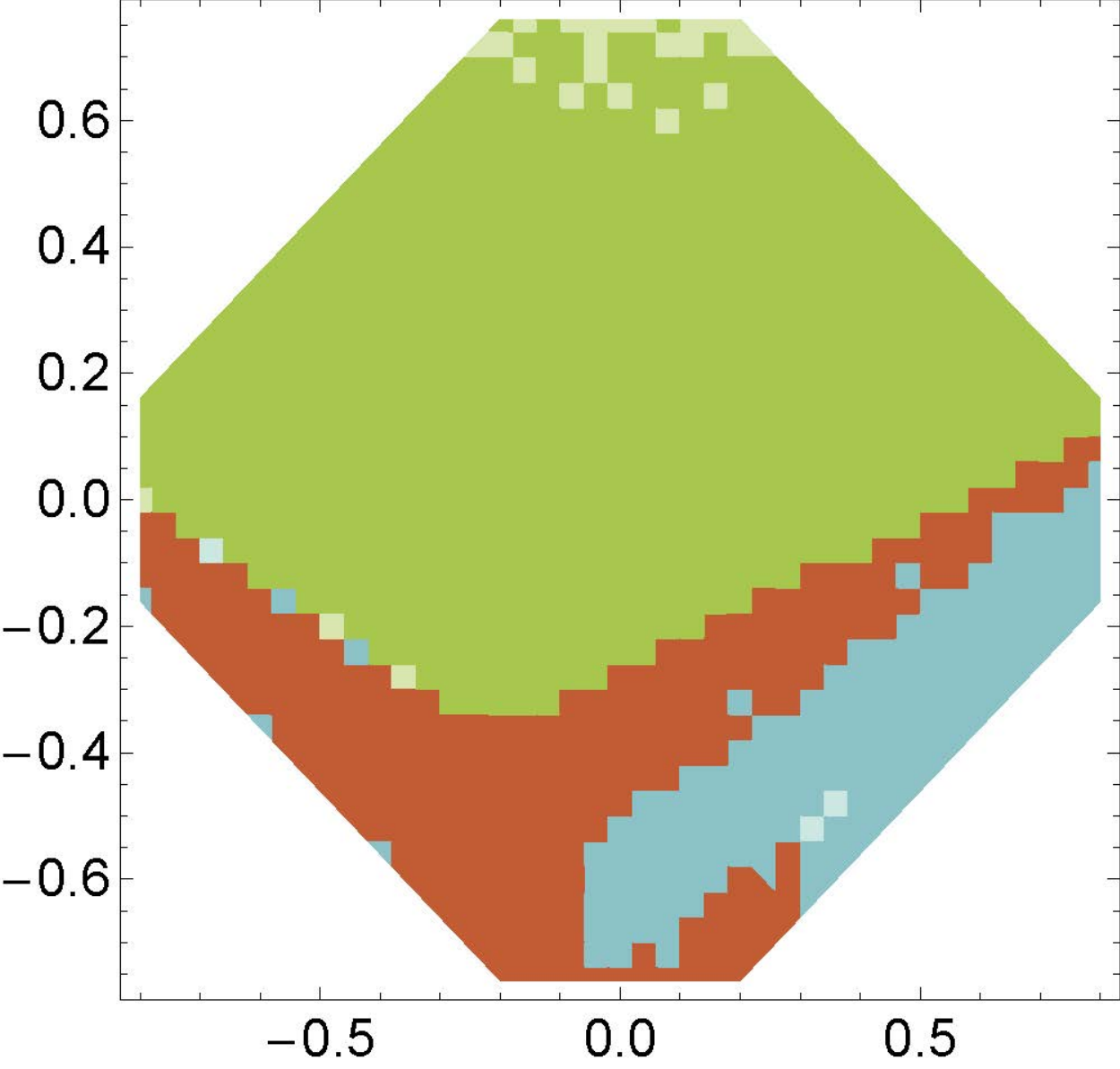}
   \hspace{10.0mm}
   \includegraphics[width=0.36\columnwidth]{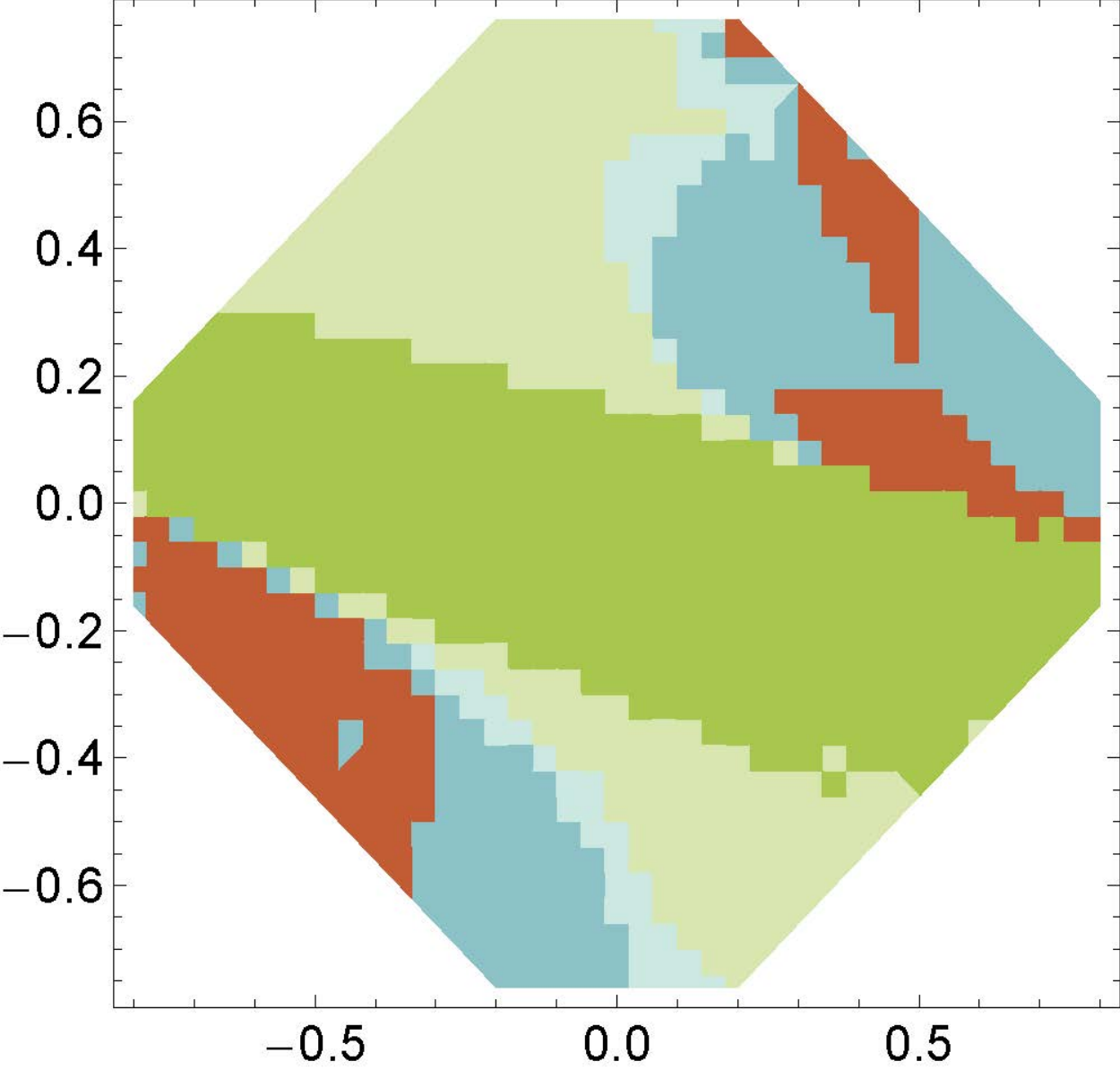}
   \end{center}
   \begin{picture}(10,10)(0,0)
   \put(125,4){\Large{(a)}}
   \put(322,4){\Large{(b)}}

   \put(193,16){\large{$a_1$}}
   \put(392,16){\large{$a_1$}}
   \put(34,170){\large{$a_2$}}
   \put(232,170){\large{$a_3$}}
   \end{picture}
   \caption{Phase diagrams showing stability of synchronized state for different combinations of parameters in driving force profiles (a) $F^t(\phi) = F_0(1+a^{(1)}\cos{\phi}+a^{(2)}\cos{2\phi})$, (b) $F^t(\phi) = F_0(1+a^{(1)}\cos{\phi}+a^{(3)}\cos{3\phi})$. The colour of each square represents the following stability: Orange represents type (i); dark blue represents type (ii); pale blue represents type (iii); pale green represents type (iv); bright green represents type (v).}
   \label{fig:harm12}
\end{figure}
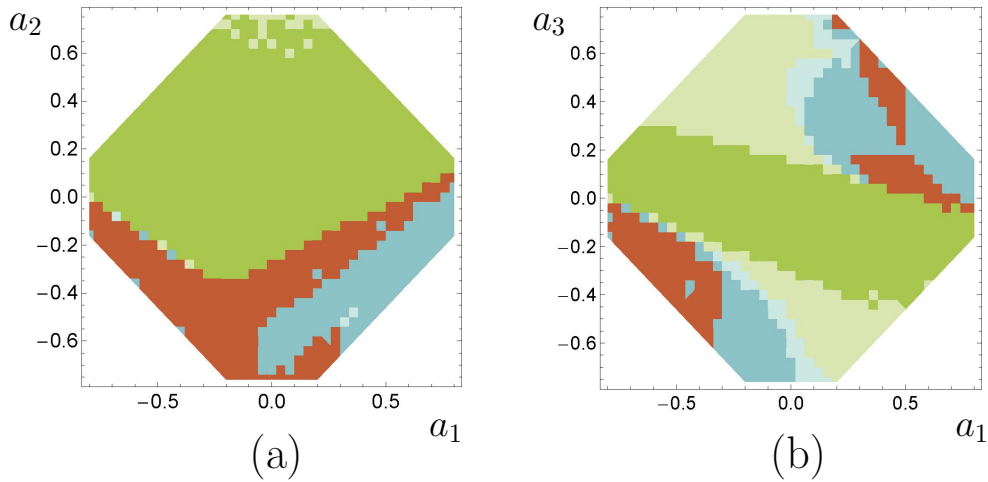
We see that there are large regions for which the synchronized state is type (i) stable.


\end{section}

\begin{section}{Conclusion}
This simple mechanical model is able to evolve into a stable synchronized state for certain choices of parameters in the driving force when the initial condition is within some region of the synchronized state. We do not need hydrodynamic interactions to achieve stable synchronization; we include hydrodynamic friction on each bead, with force free and torque free conditions and a phase dependent driving force, which can be constructed with a suitable combination of harmonic terms. For many choices of force profile, some initial conditions allow the model to evolve into the synchronized state, while other initial conditions that are very close to the synchronized state lead to an oscillating state.  There are some force profiles, including constant forcing, where there are no initial conditions that evolve into the synchronized state, and if the system starts in the synchronized state when the driving forces are equal, even a small amount of numerical noise can drive the system into an oscillating state.
There are different types of periodic behaviour for different choices of parameters in the driving force; often the phase difference oscillates about $\pi$, but sometimes the oscillations can occur close to zero with multiple peaks per cycle.

When the parameter in the driving force is different for the left and right beads, we can get periodic oscillating states about a range of values, or we can get a drifting oscillating state, where one bead has a higher average angular velocity than the other. When the coefficients have equal magnitude and opposite sign, this can lead to oscillations about the synchronized state. This frustrated synchronization is interesting when we add intrinsic noise to the driving force, because then the behaviour of $\delta$ is very similar for both opposite coefficients and equal coefficients, although the behaviour in the orientation is different in the two cases.

The nonlinear mechanics of the system make it difficult to study analytically and it is not easy to predict the parameter ranges which give stable synchronization. Our numerical results have highlighted some of the main types of behaviour of the model.

An important feature is that it is necessary to have some sort of phase dependent driving force in order to have a stable type (i) or type (ii) synchronized state. When the phase dependence is only weak, then the synchronized state is unstable, which we see from figure \ref{fig:harmstab} when the coefficient is small. The value of the coefficient at which the phase dependence becomes strong enough to give synchronization depends on the harmonic, whether we choose a positive or negative coefficient, and whether we choose sine or cosine. These latter choices are equivalent to adding a constant phase $0$, $\pi$ or $\pm \pi/2$ in the harmonic term.

This simple mechanical model shows a wide range of behaviour when we vary the parameter in the driving forces. The variety of stabilities suggests possibilities for developing run-and-tumble models, where noise can be used to to jump between regions of a phase diagram that lead to synchronization or oscillations, or jump between phase diagrams as we did in reference \cite{bg:2012}.
\end{section}

\ack
We would like to thank Nariya Uchida for fruitful discussions and the EPSRC for financial support.

\appendix
\section*{Appendix}
\setcounter{section}{1}

We note that linear stability analysis can not be used to probe the stability of the synchronization, as we cannot perform a valid Taylor expansion when $\phi=\phi_s$, where $H\cos{\phi_s}+L\sin{\phi_s}=0$. For equal driving force profiles that we linearize, $F_l^t(\phi-\delta/2)\approx F_0(\phi)-F_1(\phi)\delta/2$, $F_r^t(\phi+\delta/2)\approx F_0(\phi)+F_1(\phi)\delta/2$, if we were to Taylor expand, then the linearized expression for $\delta'$ would be
\begin{equation}
   \frac{{\rm d}\delta}{{\rm d}\phi} \approx \frac{f(\phi;F_0,F_1)}{(H\cos{\phi}+L\sin{\phi})^2}\delta,
   \label{eq:linstab}
\end{equation}
which has a singularity at $\phi=\phi_s$. The apparent singularity actually occurs at $\phi_l=\phi_s$ and at $\phi_r = \phi_s$ in the full expression, but the choice of constraining force ensures this zero in the denominator is canceled by the numerator. However, when we expand in $\delta$ and shift the singularity so that it occurs at $\phi=\phi_s$, then the numerator is no longer zero at this point. The reason we have this zero in the denominator is the following: The torque free condition (\ref{eq:forcetorquefree}) is
\begin{eqnarray}
   0&=&(\mathbf{R}_l-\mathbf{R}_b)\times \mathbf{F}_l+(\mathbf{R}_r-\mathbf{R}_b)\times \mathbf{F}_r, \nonumber \\
   &=&\big(F_l^t(-L\cos{\phi_l}+H\sin{\phi_l}+b)+F_l^n(-L\sin{\phi_l}-H\cos{\phi_l})+ \nonumber \\
   & & F_r^t(L\cos{\phi_r}-H\sin{\phi_r}-b)+F_r^n(L\sin{\phi_r}+H\cos{\phi_r}) \big)\hat{{\bf z}}. \label{eq:torquefree}
\end{eqnarray}
Along with equations (\ref{eq:Rl}-\ref{eq:Rb},\ref{eq:Rld}-\ref{eq:Rbd}), we use (\ref{eq:torquefree}) to solve for the constraining forces $F_i^n$, $i=l,r$. However, at $\phi_i=\phi_s$, $F_i^n$ is multiplied by a term which vanishes, so the torque free condition can be satisfied without specifying $F_i^n$. We over-constrain the system when we divide by zero and specify $F_i^n$ at $\phi_i=\phi_s$. Geometrically, $\phi_s$ corresponds to the phase where $\mathbf{R}_i-\mathbf{R}_b$ is parallel to $\hat{{\bf n}_i}$. 
Our numerical analysis of the full expression avoids this singularity.

\end{document}